%% file: Article.tex
\documentclass[fleqn,usenatbib]{mnras}

\usepackage{newtxtext,newtxmath}

\usepackage[T1]{fontenc}

\DeclareRobustCommand{\VAN}[3]{#2}
\let\VANthebibliography\thebibliography
\def\thebibliography{\DeclareRobustCommand{\VAN}[3]{##3}\VANthebibliography}

\usepackage{graphicx}	
\usepackage{amsmath}	
\usepackage{subfig}

\defcitealias{scott_2020_fornaxI}{Paper I}
\defcitealias{Eftekhari_2021_fornaxII}{Paper II}
\defcitealias{Romero-Gomez_2022}{Paper III}
\defcitealias{McDermid2015}{McD15}
\defcitealias{Weisz_2014_sfh}{W14}

\newcommand{\alen}{$\alpha$-enhancement }
\newcommand{\alfe}{[$\alpha$/Fe] }
\newcommand{\miles}{\href{http://miles.iac.es/}{MILES} }
\newcommand{\ppxf}{\href{https://www-astro.physics.ox.ac.uk/~mxc/software/}{pPXF} }
\newcommand{\atlas}{ATLAS$^{3D}$ }
\newcommand{\solmas}{M$_{\odot}$ }
\newcommand{\stelmas}{M$_{\star}$ }
\newcommand{\sml}{$\sim$}
\newcommand{\sami}{SAMI-Fornax }
\newcommand{\local}{Local Group }
\newcommand{\superscript}[1][]{$10^{#1}$}

\newcommand{\gls}{galaxies }

\title[The SAMI - Fornax Dwarfs Survey IV]{The SAMI - Fornax Dwarfs Survey IV. Star Formation Histories of Dwarf and Early-Type Galaxies: Insights from Full Spectral Fitting}

\author[J. Romero-G\'omez et al.]{
Romero-G\'omez, J.,$^{1,2}$\thanks{E-mail: jorgerg658@gmail.com (IAC)}
J. A. L. Aguerri$^{1,2}$,
Reynier F. Peletier$^{3}$,
Steffen Mieske$^{4}$,
\newauthor
Glenn van de Ven$^{5}$ and
Jes\'us Falc\'on-Barroso$^{1,2}$,
\\
$^{1}$Instituto de Astrofísica de Canarias, Calle Vía Láctea S/N, E-38205, La Laguna, Tenerife, Spain\\
$^{2}$Universidad de La Laguna Avda. Astrofísico Fco. Sánchez, E-38205 La Laguna, Tenerife, Spain\\
$^{3}$Kapteyn Institute, University of Groningen, Landleven 12, 9747, AD, Groningen, The Netherlands\\
$^{4}$European Southern Observatory, Alonso de Cordova 3107, 7630355 Vitacura, Santiago,Chile\\
$^{5}$Department of Astrophysics, University Vienna, Türkenschanzstrasse 17, A-1180 Wien, Austria\\
}

\date{Accepted 2023 December 6. Received 2023 November 27; in original form 2023 June 5}

\pubyear{2023}

\begin{document}
\label{firstpage}
\pagerange{\pageref{firstpage}--\pageref{lastpage}}
\maketitle

\begin{abstract}
We present a study on the star formation histories (SFHs) of galaxies covering the range \superscript[4]{} < M$_{\star}$/\solmas < \superscript[12]{}, leveraging full spectral fitting algorithms. Our sample consists of 31 dwarf galaxies from the SAMI-Fornax Survey with stellar masses between $10^{7}$-$10^{9.5} M_{\odot}$, early-type galaxies from the ATLAS$^{3D}$ project with stellar masses between $10^{10}$-$10^{12} M_{\odot}$, and dwarf galaxies that are satellites of Andromeda and the Milky Way, with $10^{4}$ < M$_{\star}$/M$_{\odot}$ < $10^{8}$. We find that galaxies from $10^{7}$-$10^{8} M_{\odot}$ exhibit the smallest star formation rates (SFRs), while the SFR increase as we move down or up in mass. In this sense, we find that some $10^{5} M_{\odot}$ galaxies have cumulative SFHs that are comparable to those of $10^{12} M_{\odot}$ galaxies. Our study shows that the evolution of giant galaxies is primarily governed by their internal properties, with timescales that do not depend on their environmental location. In contrast, dwarf galaxies below $10^{8} M_{\odot}$ can be significantly affected in dense environments, such as the inner regions of a cluster, that severely quench the galaxies before the assembly of their 50\% present-day mass. We find that, only dwarfs with stellar masses between $10^{7}$-$10^{9} M_{\odot}$ actively form stars nowadays, while less massive galaxies seem to remain unaffected by the environment due to the expulsion of most of their gas at an early stage in their evolution. Our study highlights and corroborates a critical threshold around $10^{8}-10^{9} M_{\odot}$ in galaxy evolution from previous studies, separating more massive galaxies minimally impacted by the environment from those less massive galaxies quenched by it.
\end{abstract}




\section{Introduction}\label{introduction}
One of the fascinating objects that have captivated astronomers for centuries are galaxies. They come in a variety of shapes and sizes, and their properties can vary significantly from one galaxy to another. One of the key questions in astrophysics is how galaxies form and evolve over time.
By understanding the star formation histories, one understands how galaxies form. 
Stars are the building blocks of galaxies, and their formation and evolution are intimately connected to the properties of their host galaxies. By studying the star formation histories of galaxies, we can gain insights into the physical processes that shape their evolution, such as cosmic reionization \citep{Ricotti2002, Salvadori2014, Katz2020}, feedback from supernovae \citep{Salvadori2008, Stinson2009, Sawala2010} and environmental processes \citep{Gunn1972-rampressure, Larson1980}, among others. On top of that, the shape of a SFH can be also used to study the growth of the halo at low halo masses \citep{Fitts2017, Laura2023}, or even the nature of Dark Matter, since different dark matter content predict different accretion rates \citep{Governato2015, Bozek2019}, which is then reflected in the SFH.

Ideally, to obtain the SFH of a galaxy in great detail we should study all of its stars individually. Photometry can be used to analyze the light emitted by stars, and place them in a colour-magnitude diagram (CMD) to get ages and masses \citep{Dolphin2005, Albers2019}. If deep enough, one can derive a SFH from the CMD as the amount of mass formed at different times. With the current technology, such stellar CMDs can only be constructed for galaxies within the Local Group \citep{Mateo1998, Tolstoy2009, Weisz2011, Monachesi2012}. For galaxies beyond the Local Group, one needs to analyse the integrated stellar light with spectroscopy. By analyzing the integrated light emitted by stars in a galaxy we can get the luminosity weighted mean value of its stellar population properties. To obtain SFHs, that by definition go beyond mean present-day properties, we have to compare galaxy spectra with different single stellar population models (SSP) \citep{Elodie-prugniel2007, Vazdekis2010, Verro2022}, and then use spectral fitting algorithms \citep{Capellari2004, Cid2005}. With this technique, it is possible to constrain the weights that different populations have in the galaxy spectra, and with the corresponding ages a SFH can be constructed \citep{Fahrion2021, Bidaran2022, Laura2023}.
The disadvantage of spectroscopy is that high signal-to-noise (S/N) spectra are needed, which depends on the mass and distance to the galaxy. Massive galaxies have been studied for decades, and the general consensus is that \gls with stellar masses greater than \superscript[10] \solmas formed most of their stars very fast during the early epochs of the Universe \citep{Heavens2004, Panter2007, Carnall2019}. 

Dwarf galaxies, on the other hand, are small, low-mass galaxies that are much less luminous than their larger counterparts. That is the reason why, despite being the most common type of galaxy in clusters and groups \citep{Binggeli1990, Ferguson_Binggeli_1994}, our understanding is more limited. 
First detailed studies of dwarf galaxies started with the Local Group \citep{Shapley1938b, Wilson1955, Cannon1977}, and were then extended to nearby galaxy clusters \citep[e.g., ][]{Ferguson1988, Binggeli1985, Smith2009_highalpha, Sybilska2017, Bidaran2020}. 
When looking at the stellar population ages there are star forming and quiescent dwarfs. The stellar populations of the star forming galaxies have barely been studied outside the Local Group \citep[e.g. ][]{Koleva2014}, because the stellar continuum is weak. In this paper, we will focus on the 'classical' low surface brightness dwarfs, the quiescent dwarfs, which include dwarf ellipticals (dEs)\citep{Binggeli1988} and dwarf spheroidals. These dwarfs are usually intermediate-old objects, $\sim$9 Gyr \citep{Koleva2009, Sybilska2017}, although some of them can also be young star-forming dwarfs \citep{Michielsen2008_allages, Rys2015_allages}. The age of a stellar population is an integrated (or averaged) star formation history: the older a galaxy is, the faster after the Big Bang it should have formed most of their present-day mass. Although some studies of the SFHs of dwarfs in clusters recently start appearing in the literature \citep{Fahrion2021, Bidaran2022}, our knowledge comes mostly from the \local, where detailed SFHs can be obtained.

The analysis of the observed SFHs made by \citet{Weisz2011} reveals that, independently of their morphological types, dwarfs with a \stelmas \sml\superscript[7] \solmas usually formed half of their current stellar mass \sml10 Gyr ago.
Less massive dwarfs are expected to continue the 'downsizing' \citep{Cowie1996, Thomas2005, Fontanot2009, McDermid2015}. This means that massive galaxies form fast, and as we go down in mass we expect them to form their stars more gradually. However, most of the examples we can find in the \local disagree. \citet{Weisz_2014_sfh} presented photometric SFHs of dwarfs, and argued that most less massive galaxies formed at earlier times, while higher mass galaxies formed later. This agrees with other studies that, using simulations, have also tried to disentangle the formation of dwarfs. For faint dEs, studies have shown that the quenching time, the time it takes the galaxy to form and consume the available gas, is shorter at low stellar masses (\sml\superscript[6]\solmas) than at intermediate-mass (\sml\superscript[8]-\superscript[9]\solmas) dEs \citep{Wheeler2014, Fillingham2015_108_mass_therhold, Wetzel2015}. There are many processes that can be responsible for the quenching.

When dwarf galaxies fall into a cluster environment, their star formation is believed to be suppressed through different mechanisms \citep{Gunn1972-rampressure, Larson1980, Lisker2009}, and their gas is removed by the Intra Cluster Medium, which could result in a jellyfish appearance \citep{Poggianti2016}. In the field, i.e., outside clusters or groups, however, \citet{Geha2012} found that there are no quenched dwarfs with \superscript[8] < \stelmas < \superscript[9] \solmas beyond distances of 1 Mpc from a massive host, suggesting that very few dwarf galaxies over \superscript[9] \solmas have been quenched in these regions. In addition, \citet{Geha2012} noticed that the fraction of quenched dwarfs increased with smaller distances to the host.
These environmental quenching effects have been also detected in other environments like galaxy clusters \citep{Wang2022}.
Further observations made by \citet{Fillingham2016} are consistent with these results. Studying HI surface density profiles, they found that in the local Universe, stripping was more effective for galaxies with \stelmas < \superscript[8]\solmas, which suggests a barrier where the quenching in SFHs goes from slow to fast. \citet{Gallart2015} also classified the SFHs of \local dwarfs as fast or slow, arguing that this was related to the position and/or the possible interaction of dwarfs with the Milky Way or M31. These ideas are back up by simulation studies. Using the dwarfs satellites of Milky Way-like galaxies from the AURIGA simulation \citep{Auriga2017}, \citet{Simpson2018} found that dwarfs closer to the host suffered more quenching due to ram pressure stripping \citep{Gunn1972-rampressure}. The effect was more intense at lower masses, around \sml\superscript[6]\solmas all dwarfs that are satellites or have been satellites in the past are quenched. This effect is less important the more massive the galaxy is, meaning, that the fraction of quenched galaxies decreased with increasing stellar mass. These group effects are also important in cluster like environments. Using the EAGLE simulations, \citet{Pallero2019} found that around 60\% of galaxies suffer a strong drop in their star formation rates outside the virial radius of the cluster, indicating that they were pre-processed. This means that the quenching process started before the galaxies started to fall, and then the decline got stronger from infall onward reaching the quench state inside the cluster. It exists, as well, a small fraction of galaxies that are quench even before arriving at the cluster.

Along with the environmental processes, there are other external effect capable of quenching galaxies. \citet{Weisz_2014_t70} found in their sample only two dwarfs that were quenched quickly after the Big Bang, therefore compatible with being quenched by reionization. This was supported by the observations made by \citet{Brown2014}, who proposed that reionization is the most likely cause for quenching the faintest galaxies (\stelmas \sml\superscript[4] \solmas). 
Other works based on \local objects have shown that reionization alone is not enough to quench dwarfs within a stellar mass range of \superscript[6]-\superscript[7]\solmas  \citep{Monelli2010, Hidalgo2011, Hidalgo2013}. But in terms of the process responsible for the quenching, it is clear that not only external processes are responsible for the evolution of the stellar populations, internal processes can play a role too \citep{Thomas2010}. According to simulations, supernova feedback can completely quench low mass galaxies (\superscript[5]-\superscript[7]\solmas) in time scales similar to reionization \citep{Sawala2010}. Agreeing with this, observations have also shown that supernovae feedback can quench low mass galaxies, \stelmas \sml\superscript[4]\solmas, in a short time scale \citep{Bromm2011, Gallart2021_sn_feeedback}. In addition, \citet{Benitez2015} studied a combination of reionization and supernova feedback to explain the variety of SFHs in nearby dwarfs and conclude that the relative importance of each effect depends on the mass of the object.

Ultimately, the strong suppression of star formation after reionization has been proposed as one possible explanation for the discrepancy in the missing satellites problem \citep{Simon-Geha_2007}. The issue is that there is a mismatch between the number of observed dwarfs versus the number of simulated dark halo substructures \citep{Klypin1999, Bullock2017}. While the simulated distribution of dark matter halos of comparable mass seems to match the number of observed Milky Way-sized galaxies, the number of observed dwarfs is much lower than the number of dark halos in their mass regime. 
Possible solutions to the problem include the destruction of fainter halos through reionization \citep{Quinn_1996, Alvarez_2009, Brown_2014_missing} or feedback from stellar processes \citep{Dekel_2003, Governato_2007}, but the observational tests of those proposals have not been conclusive. In addition, the effects of these processes on the galaxy's stellar mass function are expected to differ in different environments such as the field and clusters. In the end, all of these processes could leave an imprint in the formation of dwarfs.
Thus, studying their SFHs could help to solve many puzzles in understanding the evolution of dwarf galaxies.

This will be paper IV of this series of papers based on the analysis of an integral field unit (IFU) survey of dwarfs galaxies in the Fornax cluster. We study the SFH of galaxies in a broad total stellar mass range from \superscript[4] to \superscript[12]\solmas, by adding to the \sami sample maassive galaxies from the \atlas survey and less massive galaxies from the \local. The paper it is organized as follows: Section \ref{observs-reduct} summarizes the target selection, spectroscopic observations and results of previous papers of this series. In section \ref{Data_analysis} we explain the methodology used to analyse the data and to obtain the star formations histories that we present in section \ref{results}. Then we discuss the implication of our results in section \ref{discussion}, and finally, in section \ref{conclusions} we summarize our findings.
\begin{figure*}
\centering
\includegraphics[scale=0.7]{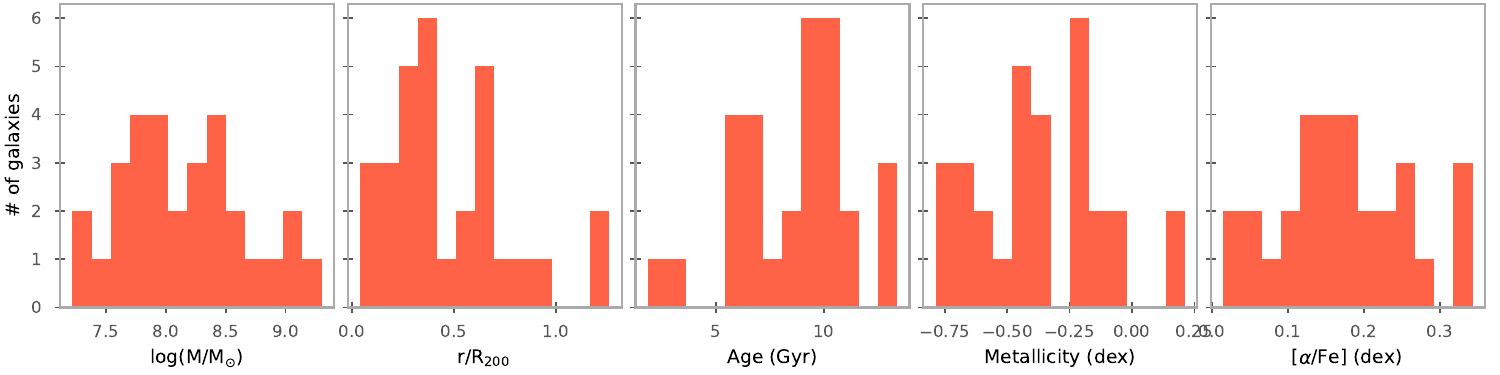}
\caption{Histograms of the distributions of different properties for our sample of dwarf galaxies in the Fornax cluster. From left to right, the panels show the stellar mass, projected cluster distance, age, metallicity and [$\alpha$/Fe]. The stellar mass is from \citetalias{Eftekhari_2021_fornaxII}, and the rest of the properties are computed in \citetalias{Romero-Gomez_2022}.}
\label{fig_intro_distribution}
\end{figure*}
The properties of the Fornax cluster used in this work are stated in the previous paragraphs. And throughout this paper, we use magnitudes in the AB system and we adopt a $\Lambda$CMD cosmology with $\Omega_{m} = 0.3$, $\Omega_{\Lambda} = 0.7$ and $H_{0}$ = 70 km s$^{-1}$ Mpc$^{-1}$.
\section{IFU spectroscopy of Fornax dwarf galaxies}\label{observs-reduct}
In this work, we use IFU observations of dwarfs in Fornax. Below we give a brief description of the selection process and spectroscopic observations. For more details on the selection of the sample, observations and data reduction we refer to the first two papers of this series \citet{scott_2020_fornaxI} \citepalias[hereinafter: ][]{scott_2020_fornaxI} and \citet{Eftekhari_2021_fornaxII} \citepalias[hereinafter: ][]{Eftekhari_2021_fornaxII}.

\subsection{Fornax dwarf sample and spectroscopic observations}\label{obs_seleciton_sec_2_1}
\subsubsection{The Fornax cluster}
Fornax is a nearby cluster of galaxies located in the southern hemisphere, $\alpha$ (J2000) $= 3^h 38^m 30^s$; $\delta$ (J2000) = -35$^\circ$27'18", and it is the second closest galaxy cluster at a distance of 20 Mpc \citep{Blakeslee2009}. It has a mean recessional velocity of 1454 km·s$^{-1}$ and velocity dispersion of 286 km·s$^{-1}$ \citep{Maddox2019}. The Fornax cluster has been the subject of numerous studies over the past few decades, which have provided valuable information about its properties. It is smaller and less massive than other nearby galaxy clusters, with a mass of $7\times10^{13} M_{\odot}$ and a 0.7 Mpc virial radius \citep{Drinkwater2001}. This makes it the densest object in the Fornax-Eridanus filament \citep{Nasonova2011}. Fornax has a sub-group of a similar mass, Fornax A \citep{Ekers1983, Maddox2019}, which is expected to fall into the main Fornax Cluster \citep{Drinkwater2001}. Photometrically the cluster has been examined in the Fornax Cluster Catalogue (FCC) \citep{Ferguson1989}, The Next Generation Fornax Survey \citep[NGFS; ][]{Eigenthaler2018_NGFS} and the Fornax Deep Survey (FDS) catalogue \citep{Venhola_2018_sample_selection}. This last catalog in \citet{Venhola_2018_sample_selection} is more complete and deeper than the others, and the newest catalog, \citet{Venhola2022}, also contains many very low surface brightness dwarfs down to a level of 27 mag arcsec$^{-2}$. This study has detected about 1000 galaxies in Fornax and Fornax A. Since, after Virgo, Fornax is the closest galaxy cluster, it is a perfect candidate to obtain the spectra of low surface brightness galaxies, like dwarfs. It represents an important laboratory for studying the formation and evolution of galaxies.
\subsubsection{Spectroscopic sample and observations}
The objects selected for the spectroscopic observations were observed at the Sydney-Australian Astronomical Observatory (AAO), using the 3.9 m Anglo-Australian Telescope (AAT) and the Multi-Object Integral-Field Spectrograph called SAMI \citep{Croom2012}. Using fibres, SAMI has 13 IFUs, each 15" in diameter, with a field of view of 1º. The observations were made in 10 pointings, with a total exposure time per field of 7 hours and a dither pattern  to ensure that the S/N was distributed uniformly on each IFU. The objects were observed in a total of 3 runs, the first in 2015 with objects selected from the FCC, and the next runs were in 2016 and 2018, using the FDS to select the objects. Apart from dEs other secondary targets were observed, like giant galaxies, UCDs or background galaxies.

The full spectroscopic sample consisted of 118 galaxies. This sample contains 56 secondary targets like UCDs, giant early-type and late-type cluster members, and background galaxies. The primary targets, 62, are classified as early-type, dE or dS0, 38 were bright enough be able to determine their internal kinematics \citepalias{Eftekhari_2021_fornaxII}. Some dwarf galaxies in this set showed prominent emission lines in their spectra. These galaxies are mainly late-type dwarf irregulars and BCDs, as classified by \citet{Venhola2019}, but there is also one quiescent dwarf that has ionized gas. These galaxies with ionised gas will not be part of the main analysis in this paper, thus, our final sample of dwarfs is composed of 31 dEs. The sample objects are located inside the virial radius of the Fornax cluster, and one is associated with the Fornax A group. The distribution of some basic properties like the stellar mass or the projected distance can be seen in Figure \ref{fig_intro_distribution}.

\subsection{Results from the previous papers in this series}
In this paper, we analyse the star formation histories of the \sami dwarf galaxy sample. In the following we briefly summarise the previous analyses made with this sample in \citetalias{scott_2020_fornaxI}, \citetalias{Eftekhari_2021_fornaxII} and \citet{Romero-Gomez_2022} \citepalias[hereinafter: ][]{Romero-Gomez_2022}.

In \citetalias{scott_2020_fornaxI}, the \sami Survey is described in terms of target selection, observation design and data reduction, and the specific angular momentum of the sample galaxies is analysed. For that purpouse, on the final spectra of the galaxies a kinematic analysis is done using the Voronoi binning techniques \citep[][]{Cappellari_Copin2003}, ensuring a minimum signal-to-noise ratio (S/N). Spatially resolved maps of the line-of-sight velocity (V) and velocity dispersion ($\sigma$) are constructed using full-spectral-fitting techniques (FSF). From these properties, the specific angular momentum ($\lambda_R$) of the galaxies is derived.
We find that dwarf galaxies exhibit significantly lower $\lambda_R$ values than more massive counterparts around $10^{10}M_{\odot}$, while only slightly higher than those seen in massive ellipticals that have almost no rotation. 
To explain those results, two scenarios are proposed: first, quiescent dwarfs with stellar masses below $10^9M_{\odot}$ could originate from low-mass spirals (\stelmas \sml $10^{10}M_{\odot}$) that lose about 90\% of their mass during their cluster infall. Alternatively, quiescent dwarfs might originate from star-forming dwarfs that fell into the cluster and were subsequently quenched by the environment. In this case, dwarf irregulars demonstrate kinematics similar to dwarf spheroidals, as ram pressure stripping has negligible impact on the stars. These unexpected results requires further studies to comprehend and verify their implications.

In \citetalias{Eftekhari_2021_fornaxII} the scaling relations on the fundamental plane (FP) and stellar mass fundamental plane are studied. The results pointed to an increase in the mass-to-light ratio of galaxies for lower masses since dwarfs with masses between $10^{7}$ and $10^{8.5}$ $M_{\odot}$ deviate slightly from the FP defined by all galaxies. Another result, based on the dynamical-to-stellar mass ratio, is that low-mass galaxies have more dark matter than brighter dwarfs and giants. In particular, dwarfs in Fornax have dark matter content comparable to the Local Group dwarfs of similar masses.

In \citetalias{Romero-Gomez_2022}, studying the galaxies as a whole and using again FSF, integrated stellar population parameters like age, metallicity and [$\alpha$/Fe] are obtained. The distribution of these and other properties of our Fornax dwarfs can be seen in Figure \ref{fig_intro_distribution}. To compare with more massive galaxies the spectra of the giant galaxies from the \atlas project are also analysed with the same methodology. The results showed that the typical [$\alpha$/Fe]-stellar mass linear relation derived for massive galaxies \citep{Worthey1992, McDermid2015} does not fit the dwarfs. Instead, a second-order polynomial is needed to fit the whole sample, displaying a U-shape relation. In terms of the environment, age and metallicity are independent for both dwarf and giants, but for dwarf galaxies, we found a linear relation between [$\alpha$/Fe] and projected distance to the cluster, with more $\alpha$-enhanced galaxies at the center \citep{Smith2009_highalpha}. Using the properties of Milky Way satellites, we were able to confirm that a relation between [$\alpha$/Fe] and the distance to the Milky Way also exists for low-mass dwarfs living in the Local Group. Additionally, we noticed that the [$\alpha$/Fe]-stellar mass U-shape derived from the Fornax dwarfs and \atlas giants, can easily be extended to the low-mass galaxies of the Local Group. These results suggests that the [$\alpha$/Fe] values of galaxies are driven by a combination of internal and external properties. Massive galaxies are barely affected by the environment, with $\alpha$-enhancement decreasing with stellar mass down to $\sim10^{9} M_{\odot}$, where we find solar-like values. For galaxies with stellar masses between $10^{7}$ to $10^{9} M_{\odot}$ the environment starts to quench galaxies as they approach a cluster or group. For even less massive galaxies, given the linear relation with projected distance, the environment seems to quickly quench galaxies, giving us high [$\alpha$/Fe] values.

\begin{figure}
\centering
\includegraphics[scale=0.54]{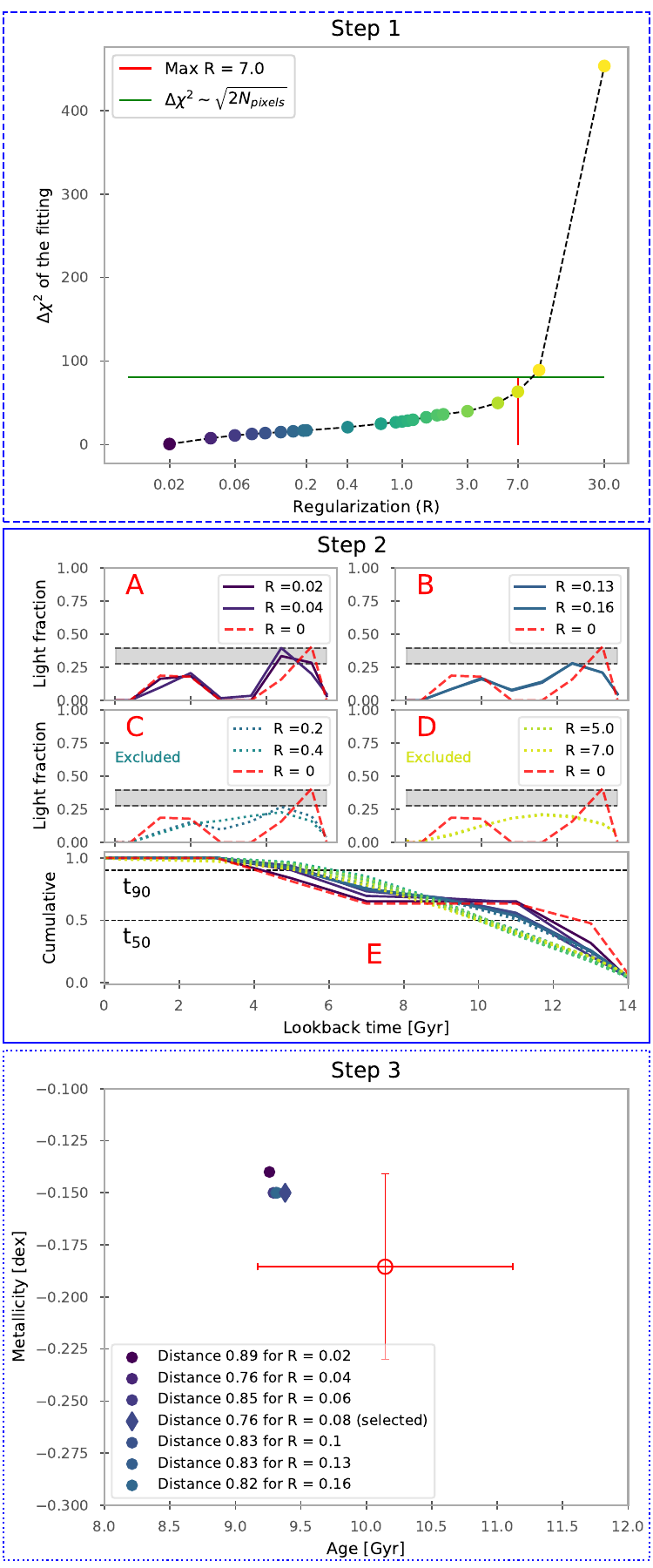}
\caption{Figure with the scheme followed during regularizaton. \emph{Top blue dashed panel, step 1}: $\chi^2$ of \href{https://www-astro.physics.ox.ac.uk/~mxc/software/}{pPXF} fitting as a function of the regularization, with each point coloured with the latter parameter. The green horizontal line represents the desired $\Delta\chi^2$, which depends only on the number of fitted pixels, and the vertical red line is the maximum tested regularization that does not surpass the desire $\Delta\chi^2$. \emph{Central blue panel, step 2}: mini panels, from A to D show the SFHs for different regularizations, marking with a dotted line those that are excluded in this step. The red dashed line represents the SFH without regularization, and the grey band in A, B, C and D indicates how much we allowed the SFHs to be smoothed. This band has been computed as the maximun peak of the SFH at R = 0, minus 1/number-of-age-models.
The mini panel E show the cumulative SFHs for all the allowed regularizaiton according to step 1, the un-regularized cumulative SFH. The two horizontal black dotted lines in this mini panel show the formation of the 50\% and 90\%. \emph{Bottom dotted blue panel, step 3}: ages and metallicities of each regularization that overlaps with the grey band in panels A to D of step 2, with a legend indicating the distance to the the stellar populations of the un-regularized fit, marked with a red cross. At this final step, if two points have the same distance, we choose the higher regularization. In the example presented in this figure we choose a regularization of 0.08.}
\label{fig_delta_chi_reg}
\end{figure}

\section{Recovering star formation histories}\label{Data_analysis}
To recover the star formation histories we first treated the spectra in exactly the same way as we did in \citetalias{Romero-Gomez_2022} to study the stellar population properties. The specific details are given in that paper, and here we provide a summary.

First, we convolved all spectra to a resolution of 2.5 \r{A} in FWHM,  which is the resolution of the spectra in the \href{http://miles.iac.es/}{MILES} stellar library \citep{Sanchez-Blazquez2006, Vazdekis2010, Falcon-Barroso2011}. These models have been used to study stellar population properties as well as the SFHs, using FSF.
With the penalized Pixel-Fitting Code (\href{https://www-astro.physics.ox.ac.uk/~mxc/software/}{pPXF}, \citeauthor{Capellari2004} \citeyear{Capellari2004} and \citeauthor{Capellari2017} \citeyear{Capellari2017}), we fitted the kinematics of each galaxy spectrum employing the \href{http://www.obs.u-bordeaux1.fr/m2a/soubiran/elodie_library.html}{ELODIE} stellar library \citep{Elodie-prugniel2007} and additive Legendre polynomials. To obtain the stellar populations and SFHs, we first have to fit the kinematics, for that we used \href{http://www.obs.u-bordeaux1.fr/m2a/soubiran/elodie_library.html}{ELODIE} models because with its resolution of 0.55 \r{A} in FWHM \href{https://www-astro.physics.ox.ac.uk/~mxc/software/}{pPXF} is capable of reaching the velocity dispersion typical of dwarf galaxies, $\sim$10 km/s \citep{Toloba2012}. Once the recessional velocity and velocity dispersion have been fixed, the stellar population properties are obtained by applying multiplicative polynomials during the fitting process. For this latter part, we decided to work with the \href{http://miles.iac.es/}{Vazdekis/MILES} models because they have a wider scope of stellar parameters that are updated frequently, even though their resolution is worst than that of SAMI.

\subsection{Star formation histories}\label{explain_regul}

After the process of obtaining the stellar population parameters with \href{https://www-astro.physics.ox.ac.uk/~mxc/software/}{pPXF} a series of weights are returned, each one corresponding to the relative contribution of the specific stellar template to the final best fit. To optimise the time performance of the fitting, we limited the ages of the models from \href{http://miles.iac.es/}{MILES} to 0.04, 0.5, 1, 3, 5, 7, 9, 11, 13, and 14 Gyr. These models also span a wide range in metallicities and a possible \alfe of 0.0 and 0.4 dex \citep[see ][ for more details]{Romero-Gomez_2022}. We tested this grid selection to ensure that the resulting stellar properties were not significantly affected by this choice. We constructed the star formation history of each galaxy, by summing up the weights returned by \ppxf over all metallicities and \alfe at a given template age, which can be seen as the fraction of the galaxy that formed at that specific time.
For our result, the \href{https://www-astro.physics.ox.ac.uk/~mxc/software/}{pPXF} fit was light-weighted, hence, we can draw the SFH of a galaxy as the light fraction formed at different epochs. 

The main problem is that the distribution of the weights from \href{https://www-astro.physics.ox.ac.uk/~mxc/software/}{pPXF} is usually discrete, meaning that it only selects a few populations from a library that contains hundreds, even if a high resolution grid is used. Consequently, the SFH will always look like a spiky function instead of a smooth distribution. For some galaxies, the SFH could really consist only of different bursts at different epochs, which in turn will look like two or three peaks. But for some others, the SFH could be a more continuous star-forming process and \ppxf will still select only a discrete number of populations. Thus, the SFH will look as if it consists only on a few bursts disperse over time, when the correct SFH would be a continuous function. We mitigate this problem using a technique called regularization \citep{Press1992}. In brief, this technique allows to identify the smoothest solution among the set of degenerate solutions that fit the data equally well \citep[see also][]{Cappellari2012_ppxf}. 
Mathematically, the smoothing of the regularization process forces two neighbouring templates to have zero derivatives, which in practice makes one population share its weight with the neighbours. 

The level of regularization, R, during the \href{https://www-astro.physics.ox.ac.uk/~mxc/software/}{pPXF} fit is controlled by its own keyword inside the code. The SFH can be smoothed as much as needed, although finding the correct regularization is a complicated process. First of all, as said before, a given SFH may not be smooth. Star formation is a stochastic process that with enough time resolution can be full of small bursts induced by different events. For that reason, while finding the correct amount of smoothness that matches the input spectra, one has to be aware of where the significant physics is: the essential information resides in the general shape of the recovered SFHs, and not in small features that can appear. This is mainly due to the degeneracy problem of the fit, since the recovered solution is not unique. Among the mathematically compatible and smooth solutions of the regularization, the weight difference between neighbours templates becomes smaller as the regularization increases, so, a flat SFH could be obtained if the regularization is set too high. 

A detailed description of the approach to address this maximum regularization searching is described in \citet{Capellari2017}. In summary, the standard procedure is as follows: first, perform an un-regularized fit adjusting the input noise (1$\sigma$ uncertainty in the galaxy spectrum) until the $\chi^2$ of the fit is \sml1, and then, raising the input regularization, while doing several tests until the difference between $\chi^2$ of the un-regularized and the current fit has increased by $\Delta\chi^2\sim\sqrt{2N}$, where $N$ is the number of fitted pixels. In the top panel of Figure \ref{fig_delta_chi_reg}, which represent the step 1, it can be seen how $\Delta\chi^2$ changes with the regularization parameter.
For each galaxy in our sample we tested different regularization values, and chose as possible regularizations those below the $\Delta\chi^2\sim\sqrt{2N}$ level,  which can be seen as an horizontal line in the top panel of Fig. \ref{fig_delta_chi_reg}. With this recipe, the smoothest SFH possible is recovered using the maximum regularization among the allowed values.

Although it is usual to use as the final SFH that of the maximum regularization \citep{McDermid2015, Fahrion2021}, we notice that for some of our dwarfs the smoothest SFH solution, though it is mathematically compatible with the data, was close to a straight line and the weights of the main populations contributing had melted away. In the central panel D of Fig. \ref{fig_delta_chi_reg} we show how the two main populations contributing to the un-regularized fit of a galaxy are replaced by an almost constant star formation rate at the maximum regularization allowed. 
In addition, the stellar population parameters of this featureless SFH were slightly different from the un-regularized results \citepalias{Romero-Gomez_2022}. For this reason, we devise a second step to the regularization procedure to ensure that the SFH is not smooth down to a straight line and at the same time the stellar population parameters are as close as possible to the un-regularized results.
So after having found the maximum allowed regularizatfion for one galaxy, we look for the maximum weight of the un-regularized fit, which defines us a top value for the SFH. While we increase the regularization, this top weight will lose its relative importance with the other populations.
We decided that this top level will only be allowed to smooth down 1/number-of-age-models, thus, creating a band like the one represented in the central panels A, B, C or D of Fig. \ref{fig_delta_chi_reg}.
We tested this smoothed number and choose it to be inversely proportional to the number of models because the more models you use, the more smooth the SFH is at R = 0. Thus, the more age models the less you need to smooth.
For example in the central panel D of Fig. \ref{fig_delta_chi_reg} it can be seen that, when the regularization is high, the SFH has been smoothed to much, the maximum allowed in that case is R=7, and the two populations from the un-regularized SFH are not distinguishable. For this reason, we select only regularization values whose SFH maximum are inside the grey band, like those in the central panels A or B, to ensure that they have not been smoothed down too much.
We want to remark that this process is just to define how smooth we want the SFHs to look, because any of the regularization values below the maximum level are mathematically compatible. As an example, in the central-bottom panel E of Fig. \ref{fig_delta_chi_reg}, we show the various possibilities for the times at which 50\% and 90\% of the current mass were formed and they are all similar for the allowed regularization values. The differences are of the same magnitude as the typical age error for these galaxies. Finally, as a fine tunning, with the selection of regularization values with a SFH overlapping with the grey band we look at the distance to the un-regularized fit in the age-metallicity plane, and chose the point for which the distance is minimal (see bottom panel of Fig. \ref{fig_delta_chi_reg}).
The [$\alpha$/Fe] parameter was not considered, since MILES only provides two possible values, 0.0 and 0.4 dex. This means that the regularization divides the weights 50/50 and the $\alpha$-enhancement was always close to or equal to 0.2 dex in all cases. Usually, the minimum distance is expected to be related to the lowest non-zero value of regularization, however, this was not the case for most galaxies since this is a highly non-linear process. Finally, once the regularization has been chosen, we fix this parameter and using the residuals of the fit we randomly perturb the best fit from \href{https://www-astro.physics.ox.ac.uk/~mxc/software/}{pPXF} using Monte-Carlo (MC) simulations and computing the uncertainties of the weights.

\subsection{Computing cumulative formation times}\label{formation_times}
Based on the age grid where a given SFH is distributed, as illustrated in Figure \ref{fig_delta_chi_reg}, the cumulative sum of the weights would reveal the time at which different mass fractions were formed. For example, the time at which 20\% of the mass was formed, t$_{20}$.
Our age grid consists of models with ages between 0.04 and 14 Gyr, so in order to have a more continuous function we interpolated the  cumulative SFHs, to extract them with a finer age grid, of ages from 0 to 14 Gyr equally spaced by steps of 0.1 Gyr. We obtained the aforementioned t$_{20}$ from the finer grid. Since the process could introduce noise and affect the calculations, we made use of the weights errors of each SFH to run MC simulations. Finally, the formation times and their uncertainties are computed as the mean and standard deviation values of the distribution, respectively.

\section{Results}\label{results}
This section presents the resulting SFHs computed with \href{https://www-astro.physics.ox.ac.uk/~mxc/software/}{pPXF} and the \href{http://miles.iac.es/}{MILES} library. 
As stated before, we limit this analysis to dwarf galaxies without emission lines, which leaves us with a sample of 31 dEs covering a stellar masses between \superscript[7]{} and \superscript[9]{} \solmas.
\begin{figure*}
\centering
\includegraphics[scale=0.87]{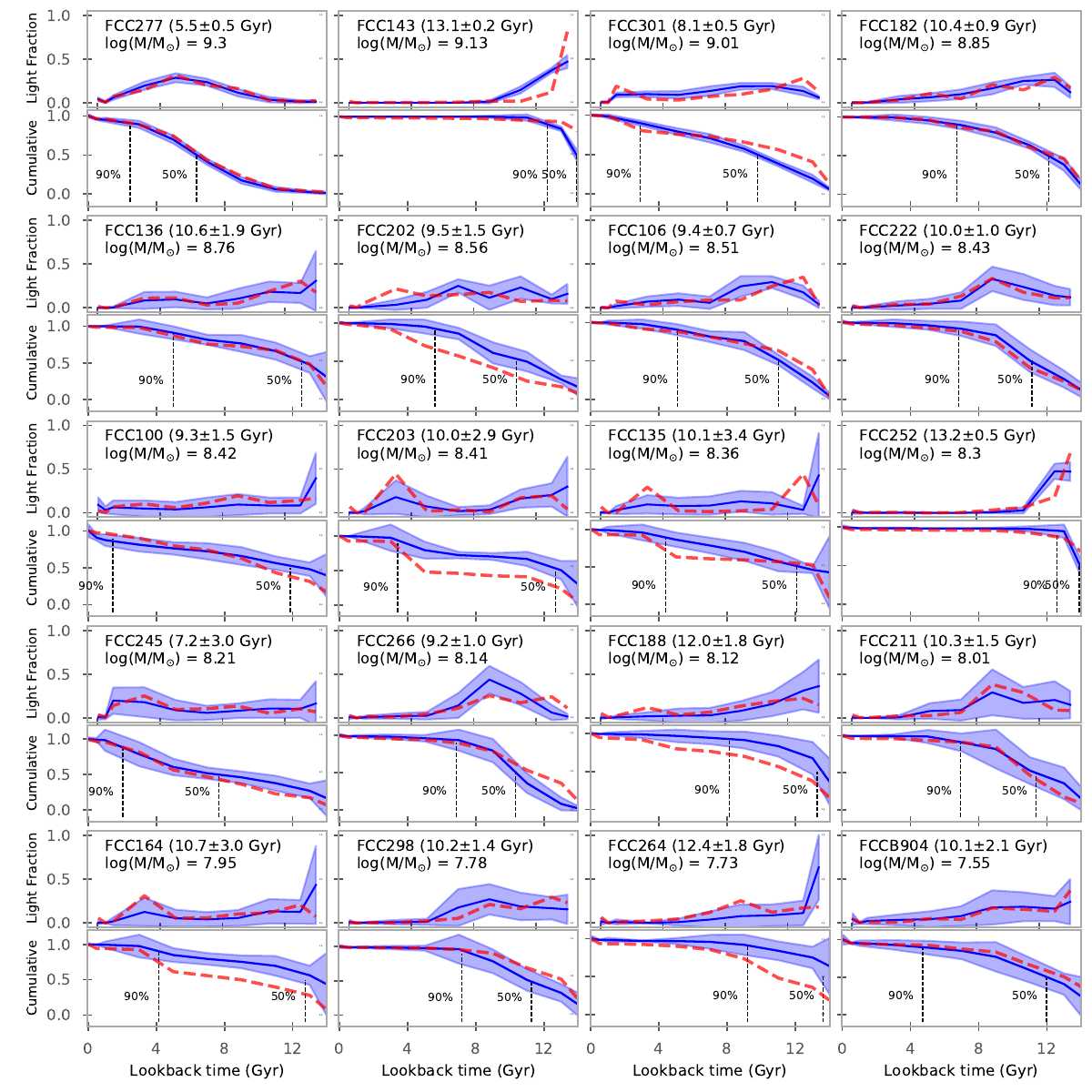}
\caption{Star formation histories of dEs in our SAMI-Fornax sample. The galaxies are ordered according to their stellar mass from left to right and from top to bottom. Only galaxies with S/N higher than 15 are included here. For each galaxy, we show two panels, one with the light fraction and the other with the cumulative fraction, both as a function of the lookback time (Gyr). In each panel, the mean SFH from the MC simulations is shown with a blue line, and the errors are the light-blue shadow. For comparison of the smoothing process, the red dashed line represent the unregularized SFH. We also included the name of each galaxy, the mean light-weighted age obtained from the SFH, and the corresponding stellar mass. The curves with the cumulative fraction incorporate two vertical dashed black lines indicating the formation time of the 50\% and 90\% of the present-day stellar mass.}
\label{fig_sfh_sn_26}
\end{figure*}
\subsection{Star formation histories of dwarf galaxies}
In Fig. \ref{fig_sfh_sn_26} we present the SFHs of SAMI-Fornax dEs.
We also included vertical lines indicating two characteristic formation times, t$_{50}$ and t$_{90}$. These are useful and interesting parameters that indicate the time when the 50\% and 90\% of the mass was formed. It is usually considered that $t_{50}$ is related to the formation time of the galaxy halo \citep{Tojeiro2017_t50_halo}, while $t_{90}$ is considered as a proxy of the quenching time of the galaxy, when the star formation has entirely ceased \citep[see][]{Ferre2018}.

In our sample of dEs, 80\% of them formed half of their present-day mass more than 8 Gyr ago, with a mean $t_{50}$ of 11 $\pm$ 3 Gyr. But only a few of them built this percentage of mass in less than 2 Gyr since the Big Bang. On the other hand, we notice that only three of our dwarfs are compatible with being already quenched 12 Gyr ago, and by 4 Gyr ago only half of our SAMI-Fornax sample put a stop to their star-forming activity. The mean $t_{90}$ of our sample is 5 $\pm$ 3 Gyr. These numbers are not surprising, since the SFH is just the time distribution of the different populations that conforms a galaxy, and our sample is composed mostly of intermediate-old ages (see \citetalias{Romero-Gomez_2022}). For most of our galaxies, the star-forming process took several Gyr to increase from 50\% to 90\% of their current stellar mass, as can be seen in Fig. \ref{fig_sfh_sn_26}. Note that this process only lasted for about $\sim$3 Gyr for 10\% of our sample. This is an indication of how fast these galaxies were formed.

\subsection{Comparison with the star formation histories of massive galaxies}
As in our previous work, we have also re-analysed the data from the \href{https://www-astro.physics.ox.ac.uk/atlas3d/}{ATLAS$^{3D}$ Project} \citep{Cappellari2011-atlas3d}, to be able to compare our results with a sample of massive galaxies. In \citetalias{Romero-Gomez_2022} we reported the ages, metallicities and $\alpha$-enhancement of these galaxies. Here we present the complementary SFHs, using the exact same methodology explained in section \ref{Data_analysis}. The ATLAS$^{3D}$ sample is composed of early-type galaxies (ETGs), with a stellar mass between $10^{10}$ and $10^{12} M_{\odot}$. Some of these galaxies belong to the Virgo cluster, and some of them are in the field.

As we did for our SAMI-Fornax dwarfs, for every object in the ATLAS$^{3D}$ sample we have derived the SFH and the corresponding formation times. To test our results we have also compared the SFHs and the time at which 50\% of the mass was formed with the results published in \citet{McDermid2015}\citepalias[hereafter:][]{McDermid2015} (see appendix \ref{appendix_atlas} for more details).
In \citetalias{McDermid2015}, they notice some issues during the spectral analysis of some galaxies for not having enough S/N, emission lines or some other problems. We have excluded the galaxies that display these effects as well. 
\begin{figure*}
\centering
\includegraphics[scale=0.88]{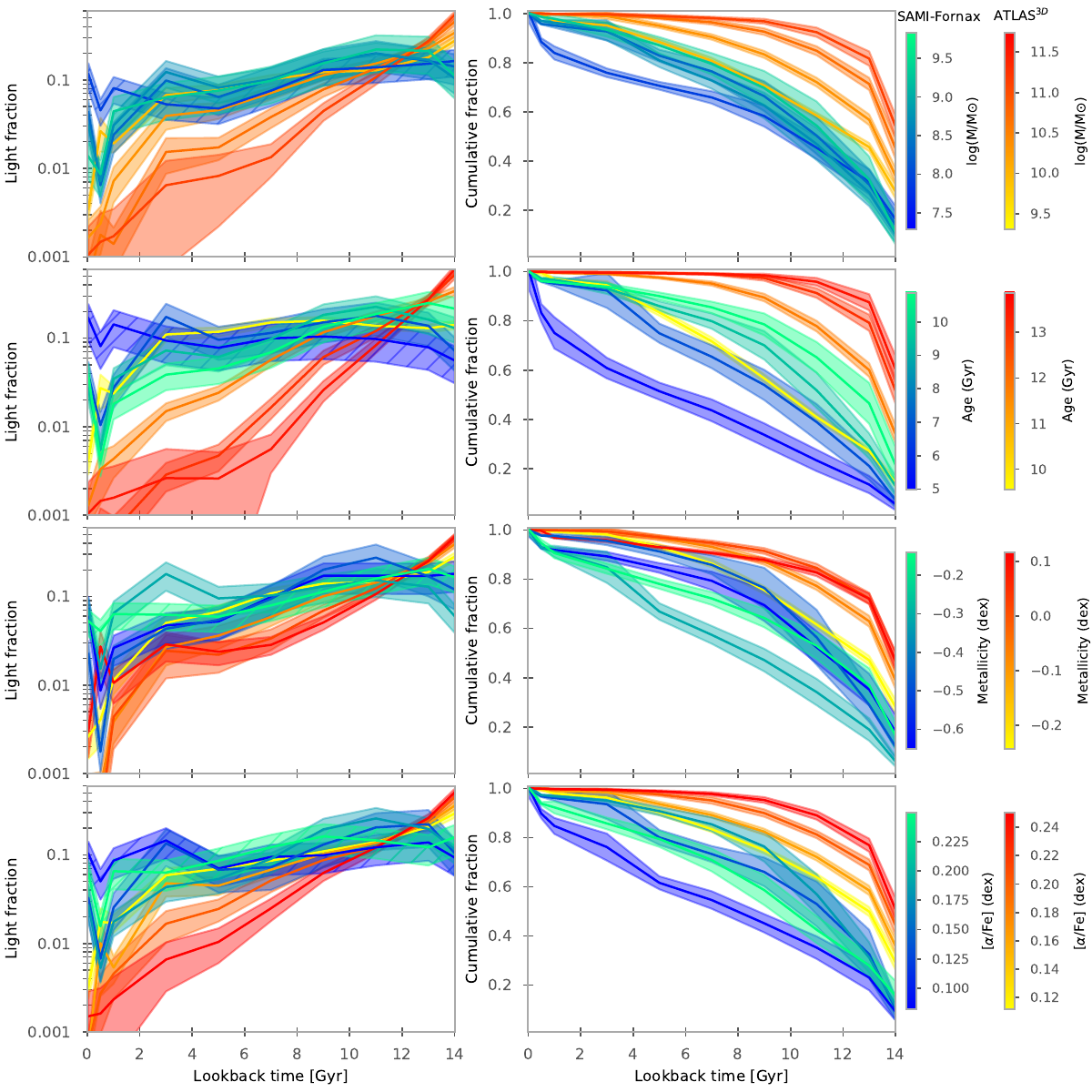}
\caption{SFHs (left column) and cumulative fraction (right column) of SAMI-Fornax and ATLAS$^{3D}$ samples color coded as function of different galaxy properties. We grouped the galaxies into four equally spaced bins, for the different properties displayed in the colour bars. For each bin, the SFH is the mean of the SFHs of the galaxies inside the bin, and the error is the standard deviation. These bins are colour-coded according to the property in the colour bars with the mean value of galaxies inside that stellar-mass bin. From the top row to the bottom, the properties displayed are logarithmic stellar mass, age, metallicity and [$\alpha$/Fe]. We separate with distinct colour maps the SFHs from bins containing dEs and ETG. With blue to green colours we represent the dwarfs in our SAMI-Fornax sample, and from yellow to red the giant galaxies from ATLAS$^{3D}$.}
\label{fig_SFH_with_atlas_2bars}
\end{figure*}
With the combination of the SAMI-Fornax and ATLAS$^{3D}$ samples, we have the SFHs of a total of 260 galaxies with stellar masses between \superscript[7]-\superscript[12]\solmas. To be able to compare that many galaxies, we grouped them into bins depending on different properties: stellar mass, age, metallicity and \alfe. We used four equally spaced bins for the dwarfs and another four for the giants. The final cumulative SFHs can also be seen in Fig. \ref{fig_SFH_with_atlas_2bars} as a function of different internal properties of the galaxy.

As expected, the shape of the SFHs is tightly related to the stellar mass, as shown in the top panel of Fig. \ref{fig_SFH_with_atlas_2bars}. More massive galaxies formed at earlier epochs, while less massive galaxies formed slower and took a longer period of time. This trend is visible for both giant and dwarf galaxies, although it is less pronounced for the latter. The most massive galaxies of the ATLAS$^{3D}$ sample formed built up to 90\% of their present-day mass within about 2 Gyrs after the Big Bang, while the less massive dwarfs have been forming stars until recently.
For intermediate masses, 10$^{9}$ < \stelmas < 10$^{10}$ M$_{\odot}$, we notice that the cumulative SFHs of both the lowest mass galaxies from the \atlas sample and the more massive dwarfs from the \sami follow a similar path and it would be difficult to distinguish them only with the information of the cumulative SFH. 
This is expected since we are always looking at early-types galaxies, although the classification of these objects in \atlas \citet{Cappellari2011-atlas3d} and SAMI-Fornax \citet{Venhola_2018_sample_selection, Venhola2019} is different, probably due to the fact that each group uses a different definition for classifying a galaxy as dwarf. But since, independently of the morphology, we see that the star-formation rate is faster, the more massive the galaxy is. This star-formation rate that goes from massive galaxies to intermediate-mass giants, gradually connects to massive dwarfs and then continues to the less massive dwarfs.

In the second panel of Fig. \ref{fig_SFH_with_atlas_2bars} we can see how older galaxies are more massive than younger galaxies. This is in agreement with other works of the literature where usually a relation between age and stellar mass is found \citep{Sybilska2017, Romero-Gomez_2022}. In the third panel of Fig. \ref{fig_SFH_with_atlas_2bars} we show the relation of the SFHs with metallicity. The metallicity-stellar-mass relation is a well know relation for galaxies \citep{Kirby2013}. Since  the cumulative SFHs are related to stellar mass, it is expected they are related to the metallicity too. We can see that the \atlas galaxies follow this trend, while for dwarfs the trend is somewhat less clear.
For both of these last two properties, we see again that giants and dwarfs with similar ages or metallicities have indistinguishable evolutionary paths.

In the fourth row of Fig. \ref{fig_SFH_with_atlas_2bars} we can see the SFHs according to their \alen. Since we used the \miles models, our range in \alfe is limited, only from 0.0 to 0.4 dex. Both dwarfs and giants cover the same values and there is no segregation between the different morphological types in terms of [$\alpha$/Fe] range. However, for the \atlas galaxies we see clearly a relation between the cumulative SFHs and the \alen values. Giants with higher enhancements formed faster a long time ago, and those with lower \alfe have been building up more slowly, until a few Gyr ago \citepalias{Romero-Gomez_2022}. This is associated with the fact that there is a \alfe-stellar-mass linear relation for giant galaxies, but this is not the case for the dwarfs. We see the cumulative SFH of low \alfe below the curve of high \alfe, but we also see another low \alfe curve following a similar path to those of higher \alen. This is because for the mass regime of the dwarfs, the \alfe values are not only a function of the stellar mass, but also of the environment \citepalias{Romero-Gomez_2022}. In general, as it will be discuss later in section \ref{discus_timescales}, the [$\alpha$-Fe] values could be used as an interpretation of the timescales of star formation in a galaxy.

\begin{figure}
\centering
\includegraphics[scale=0.67]{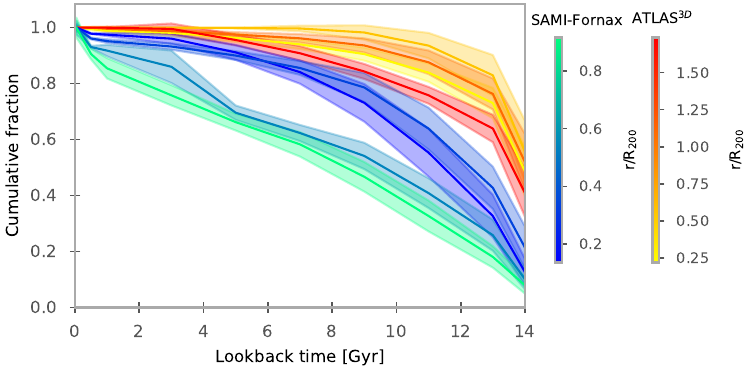}
\caption{Cumulative SFHs of SAMI-Fornax and ATLAS$^{3D}$ samples colour coded as a function of their distance to the cluster. We grouped the galaxies according to their r/R$_{200}$ into 4 bins. For the giants, we only use here those that fill our definition of cluster members, and the Virgo membership definition from \citet{Cappellari2011-atlas3d}. The colour codes are the same as in Fig. \ref{fig_SFH_with_atlas_2bars}.}
\label{fig_SFH_with_atlas_2bars_dist}
\end{figure}

For the mass regime of the dwarfs, the formation timescales can be affected by the environment \citep{Smith2009_highalpha}. In Fig. \ref{fig_SFH_with_atlas_2bars_dist} we show the cumulative SFHs in relation to the galaxy projected distance to the center of the cluster. Based on the mass range of the giant sample, some of them could also be affected by the environment, but if we bin them by distance, the \atlas galaxies inside the virial radius of Virgo do not show any meaningful signs of the SFHs being strongly affected by their present-day environment. Only when we move farther away than the virial radius, we see that giant galaxies have slower cumulative SFHs. On the other hand, the dwarfs in Fornax, all of them inside or close to the virial radius of the cluster, show a clear dependence on the present-day environment distance. The relation represents how the environmental quenching of the star formation, for those galaxies closer to the center of the cluster, start to be effective a few Gyrs after the Big Bang.

\subsection{Formation times}\label{results_formation_times}
We now analyze the formation times of the galaxies, defined as the parameters t$_{50}$ and t$_{90}$, i.e., the times that 50 and 90\% of the stars were formed, respectively. In order to have a sample of galaxies covering also the lower end of the mass spectrum, we need to include the t$_{50}$ and t$_{90}$ values of the galaxies in the Local Group.
For less massive galaxies is a challenge to get an accurate spectroscopic observation, given their low surface brightness, and when they are close enough to be easily observed, like in our \local, it is impossible to get one spectrum that covers the whole galaxy. For that reason we have relied on the photometric SFHs from \citet{Weisz_2014_sfh}\citepalias[hereafter:][]{Weisz_2014_sfh}. Their sample is composed of 40 \local dwarfs, satellites from both the Milky Way and Andromeda. They obtained the SFHs by analyzing the colour-magnitude diagram (CMD) and using single stellar populations to assemble an artificial CMD to fit the observed populations \citepalias[see][for more details]{Weisz_2014_sfh}. Since their methodology is different from ours the numbers and timescales of their SFHs will not be directly comparable to ours, but they should show similar results. For that reason, what we are interested in, is the overall shape of the cumulative SFHs and the relation with the different galaxy properties. 

\begin{figure}
\centering
\includegraphics[scale=0.67]{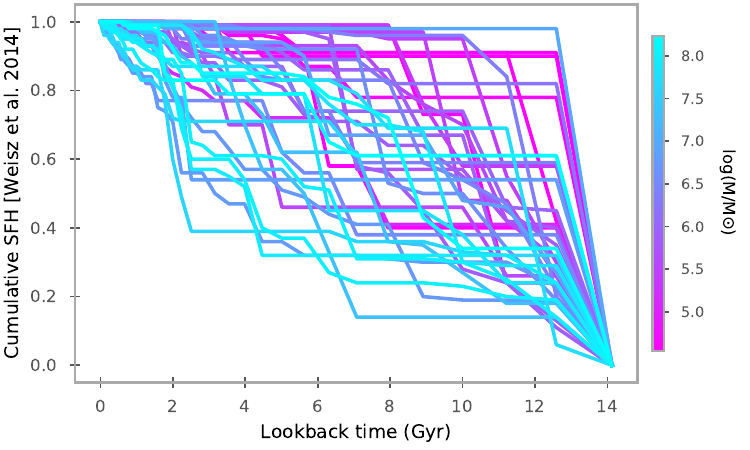}
\caption{Cumulative SFHs as a function of the lookback time (Gyr) directly as given by \citetalias{Weisz_2014_sfh} (see appendix \ref{appendix_Local_weisz}). For each galaxy, we used the corresponding colour code of the stellar-mass also published in \citetalias{Weisz_2014_sfh}. We applied a colour map where the less massive galaxies are purple, and then the colour gradually goes cyan for the most massive galaxies. The galaxies represented here are all dSph from the Local Group, including both Milky Way and Andromeda satellites. This sample is included to consider dwarfs down to a stellar mass o \superscript[4]{} \solmas.}
\label{fig_weisz_sfh_lg}
\end{figure}

In Fig. \ref{fig_weisz_sfh_lg} we show the cumulative SFHs coloured by the stellar-mass of each galaxy, the curves are directly taken from \citetalias{Weisz_2014_sfh}. Here we have selected only the dSph from their sample, avoiding the dIrr and dTrans to have a similar filter selection to our dwarfs from Fornax. In Fig. \ref{fig_SFH_with_atlas_2bars}, where we consider galaxies in the mass range \superscript[7]{}-\superscript[12]{} \solmas, we identified the trend that for less massive galaxies (in the Fornax cluster) the star-formation rates were smaller. However, when considering the \local dwarf galaxies that cover the stellar mass range from \superscript[4]{}-\superscript[8]{} \solmas, we see an opposite trend.
Those \local galaxies that have a stellar-mass in the same range as our dwarf galaxies from the \sami sample, follow a similar evolutionary path. The less massive galaxies, with masses below \superscript[7] \solmas, have SFH curves more analogous to those of giant galaxies. To quantify this effect, we now consider the formation times t$_{50}$ and t$_{90}$.

\begin{figure}
\centering
\includegraphics[scale=0.8]{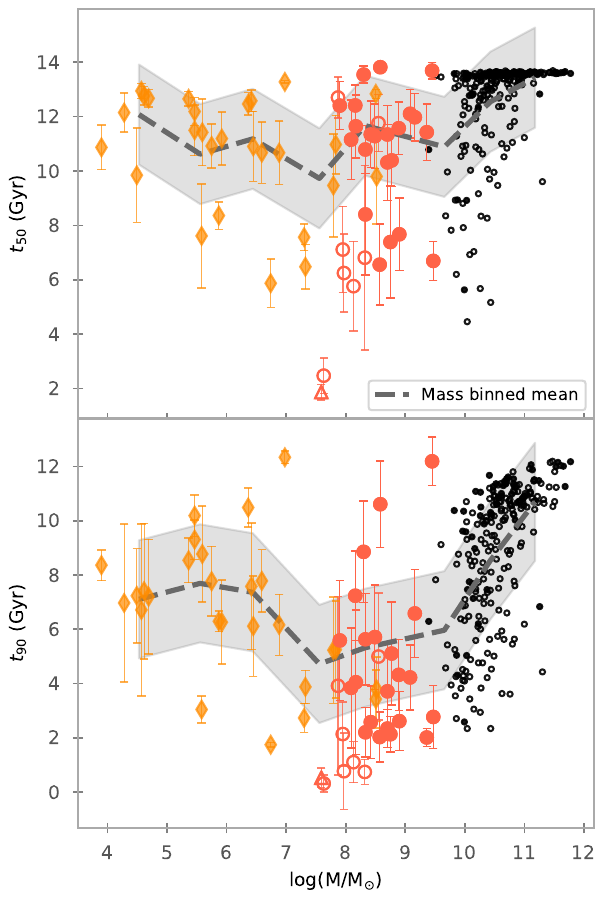}
\caption{Derived formation times of the galaxies as a function of their stellar-mass. The panels show, from top to bottom, t$_{50}$ and t$_{90}$, respectively. The giants from the ATLAS$^{3D}$ sample that we defined as cluster galaxies are the filled black dots, and the empty black dots are the field ETGs. The red circles are the dEs from our SAMI-Fornax sample, and the triangles are the dEs located in the Fornax A sub-group. These red symbols are filled if the galaxy has S/N > 15, otherwise is empty. The orange diamonds are the Local Group galaxies from \citetalias{Weisz_2014_sfh}. In both panels we added a dashed grey line representing the mean formation time for different mass bins, and a shadow with the mean error. We used bins of 1 log(M/\solmas) to cover the whole range of stellar masses.}
\label{fig_times_vs_mass}
\end{figure}
For a first comparison of the formation times, we used the SFHs from \citetalias{Weisz_2014_sfh} to compute the t$_{50}$ and t$_{90}$ times. For that, as in our galaxies, we interpolated the cumulative curves with a function that was then used to extract the SFHs with a finer grid. The formation times are obtained as the closest value from the grid to the desired mass fraction. With the errors provided by \citetalias{Weisz_2014_sfh}, we repeat the process running MC simulations. Finally, we obtained the formation times as the mean value and the standard deviation of the distributions are used as errors. To test these estimations, in the appendix \ref{appendix_Local_weisz} we show how the SFHs of \citetalias{Weisz_2014_sfh} looked after this interpolation process. Also in a following paper they used the SFHs to compute the t$_{70}$ \citep{Weisz_2014_t70}, and a comparison shows that we obtained the same formation times (see Fig. \ref{fig_compare_times_weisz_apx}).

In Fig. \ref{fig_times_vs_mass} we present the t$_{50}$ and t$_{90}$ as a function of the stellar-mass, for all the samples. We divided the \atlas sample into cluster and non-cluster according to the density parameter from \citet{Cappellari2011-density}. We defined galaxies with a local mean surface density higher than log($\Sigma_{10}$) > 0.6 Mpc$^{-2}$ as cluster members.
We see that most of the \local galaxies took less time than our Fornax dwarfs to reach 90\% of their mass, but they also took more time than massive galaxies, which is confirmed by the statistics of each sample. For the t$_{90}$, the quartiles (25\%, 75\%)  of the \local and \sami dwarfs are (5, 8) and (1, 5) Gyr, respectively, while for the giants go all the way up to (7, 11) Gyr. The times for the less massive dwarfs in the \local are therefore in between the \atlas and \sami samples. To make this more visible, in Fig. \ref{fig_times_vs_mass} we also show with a dashed line the mean formation time for different mass bins of 1 log(M/\solmas). The combined sample, covering stellar masses from \superscript[4]{} to \superscript[12]{} \solmas, defines a U-shape, with the minimum around log(M/\solmas)=8. Only in galaxies between 10$^7$ to 10$^{10}$ M$_{\odot}$ there are objects that have been forming mass until recently. For the formation time of 50\% of the galaxy mass, we see the same effects, although somewhat diluted. The \local objects have times closer to our dwarfs in Fornax, with mean values between 10-11 Gyr, while most of the massive galaxies had already formed 50\% of their mass 13 Gyr ago.

\section{Discussion}\label{discussion}
In this work, we have analyzed the spectroscopic SFHs of a sample of dwarf galaxies. We have compared them to more massive galaxies from the \atlas project, and other low-mass dwarfs from the Local Group. The results point out two key factors shaping the SFHs, the stellar mass and the projected distance to the cluster center. These two quantities are representative of internal and external processes, respectively. In the following we consider now how these two aspects affect other properties of the galaxy.

For the next figures we want to highlight that the individual values of the represented properties were obtained with different methodologies, which could introduce systematic effects into the comparison. With this word of caution in mind, we want to study the general behaviour of different galaxy properties and their relation with internal and external properties, with these figures about the projected distance-stellar mass plane.

\subsection{From giants to dwarfs}
\begin{figure*}
\centering
\includegraphics[scale=1.2]{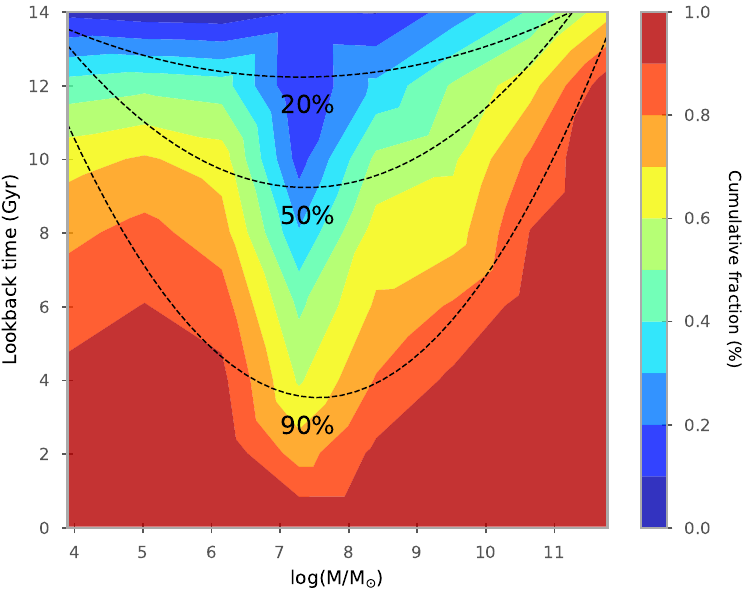}
\caption{Color map of the cumulative SFHs as a function of the galaxy mass. The vertical axis represents the lookback time in Gyr, and the colours are the cumulative fraction of stellar-mass formed at different times and at different stellar-masses. The dashed black lines represent a second-degree fit to the formation times, as shown in Fig. \ref{fig_times_vs_mass}, as a function of the stellar-mass.}
\label{fig_2Dcolor_map_mass}
\end{figure*}

Figure \ref{fig_2Dcolor_map_mass} shows a 2D color map interpolation of the SFHs that represents a complete picture of how the SFHs evolve with stellar-mass. Since we used the same finer age grid to derive the formation times for all the samples described in this paper, we can put all the cumulative SFHs together creating a uniformly distributed 2D array. Here we ordered the galaxies as a function of stellar mass, having in total a stellar mass range from 10$^4$ to 10$^{12}$ M$_{\odot}$. To create a continuous 2D image of SFHs as a function of the stellar-mass, we used the \href{https://docs.scipy.org/doc/scipy/reference/generated/scipy.spatial.KDTree.html}{Scipy.KDTree} algorithm \citep{KDTree1999} to look up the nearest neighbours of every point with which we created the smooth map in Fig. \ref{fig_2Dcolor_map_mass}.

\begin{figure}
\centering
\includegraphics[scale=0.8]{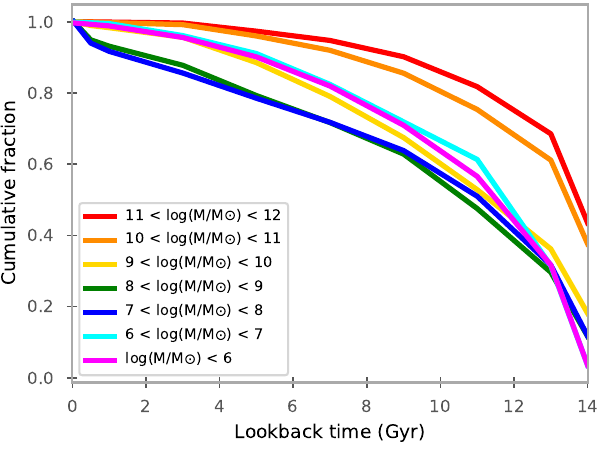}
\caption{Cumulative SFHs of galaxies for of different stellar masses. We grouped the galaxies according to their stellar mass, and plot the mean SFH of each bin. In blue we show the mean SFH for those galaxies with M$_{\star}$ < 10$^{6}$ M$_{\odot}$, green for galaxies with 10$^{6}$ < M$_{\star}$ < 10$^{8}$ M$_{\odot}$, yellow for 10$^{8}$ < M$_{\star}$ < 10$^{10}$ and red for the massive galaxies 10$^{10}$ < M$_{\star}$ < 10$^{12}$. The bins are also stated in the legend of the figure.}
\label{fig_SFH_bin_mass_all}
\end{figure}

In our previous results, we have seen how the SFH smoothly varies as we move from giants to dwarfs. This could indicate that independently of the morphology the star-forming rate is primarly related to internal properties, i.e. the stellar-mass. This is clearly shown in Fig. \ref{fig_SFH_bin_mass_all}, where we used different bins of stellar mass to group the galaxies and plot the mean SFH. We mixed here all the samples, meaning that in the same bin there could be a mix of \atlas and \sami galaxies, or cluster and Local Group dwarfs. As expected from Fig. \ref{fig_times_vs_mass}, the cumulative SFHs become more extended as we go down in mass, however, we again see how galaxies with total stellar mass < \superscript[6]M$_{\odot}$ formed faster than those galaxies with total stellar masses between 10$^7$ to 10$^{10}$ M$_{\odot}$.

For higher mass galaxies our map agrees with the results of \citet{Thomas2005} and \citetalias{McDermid2015}. On average these galaxies formed at least 50\% of their mass in just a few Gyr after the Big Bang, the more massive were quenched faster. Meanwhile, less massive objects are built more and more gradually. This effect is called 'downsizing' \citep{Cowie1996,Fontanot2009}, and it relates the galaxy mass with some key properties like the star-forming rate or the stellar population properties. However, this relation does not seem to persist at even lower masses since we detect a change in this trend around 10$^7$-10$^{8}$ M$_{\odot}$. In their study of the Local Group dwarfs, \citet{Weisz_2014_sfh} already pointed out this reversed trend, but they did not consider it to be evidence against the galaxy downsizing. They argue that this 'upsizing' is caused by the environment since most of the less massive galaxies from their sample are inside the virial radius of the Milky Way or Andromeda. This makes them vulnerable to environmental processes, caused by the circumgalactic medium in the halo, that can play a similar role as ram-pressure stripping \citep[][]{Putman2021, Akins2021-simulations}. One has to note the fact that at the low mass end, other factors such as reionization as well as internal quenching can also play a role.

Analyzing Fig. \ref{fig_times_vs_mass} and \ref{fig_2Dcolor_map_mass}, we realize that this behaviour is comparable to the [$\alpha$/Fe]-stellar mass relation we proposed in \citetalias{Romero-Gomez_2022}, which looks like a U-shape. With this in mind, we fitted a U-shape to the formation times of galaxies from Fig. \ref{fig_times_vs_mass} and overplotted them onto the colormap, conforming once again this U-shape, which agrees well with the colormap of Fig. \ref{fig_2Dcolor_map_mass}.

The change in the trend, giving rise to the U-shape, could be considered a mere bias, caused by the more massive dwarfs being located outside the virial radius of their host. However, in our previous study, we argue, analogously to \citet{Weisz_2014_sfh}, that 'upsizing' part of the U-shape is in fact due to the environment and is indeed a sign that the environmental processes are more and more effective the less massive a galaxy is. 

\begin{figure*}
\centering
\includegraphics[scale=0.78]{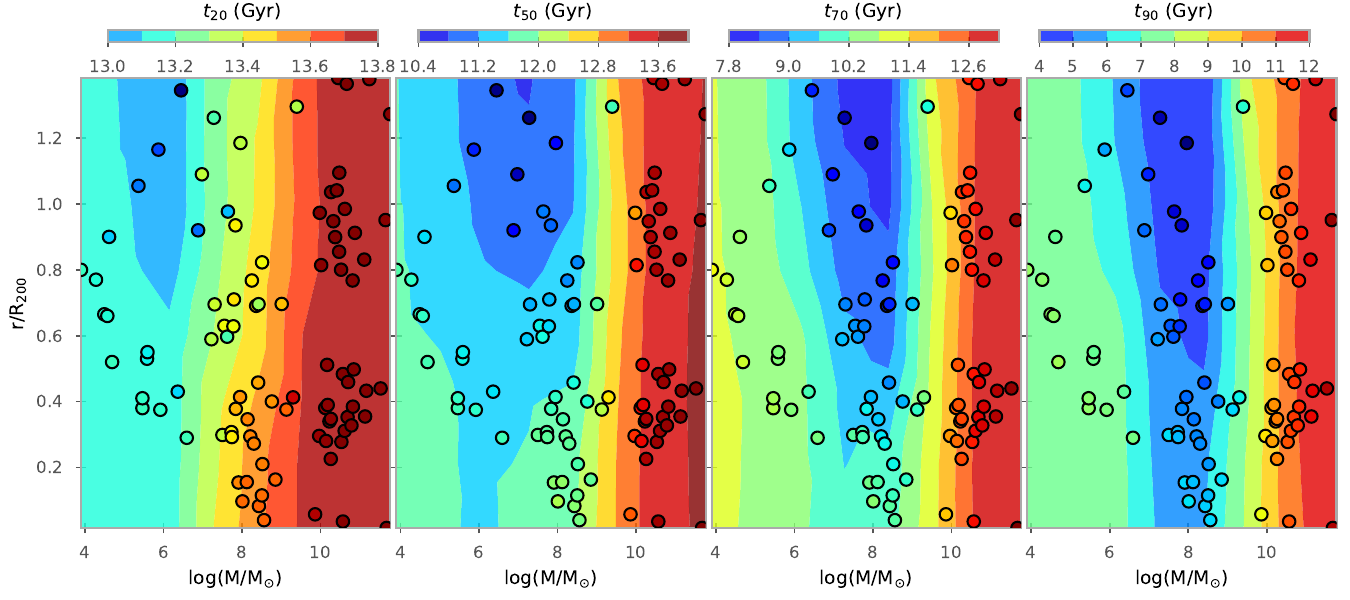}
\caption{Color map of the different formation times of the galaxies as a function of their position in the projected distance-stellar mass plane. The panels represent the time it took to form a certain percentage of the current day mass. From left to right we show t$_{20}$, t$_{50}$, t$_{70}$, t$_{90}$. The scatter points are smoothed using the LOESS technique, and then with \href{https://docs.scipy.org/doc/scipy/reference/generated/scipy.spatial.KDTree.html}{Scipy.KDTree} the background maps are extrapolated.}
\label{fig_maps_times}
\end{figure*}
\subsection{Relation between the environment and the formation of galaxies}
To further study this interplay between the internal and external properties, we plot the different formation times computed from the SFHs in a cluster/group center projected distance-stellar mass plane. We argued that the 'upsizing' part of the U-shape [$\alpha$/Fe]-stellar mass relation is due to environmental processes mainly because we found a linear relation between [$\alpha$/Fe] and the projected distance for the dwarf galaxies. This relation is present independently if they are cluster or group galaxies \citepalias{Romero-Gomez_2022}. Another reason to explain the 'upsizing' with environment is that quiescent dwarfs are mostly not found far away from their massive companions. 
\begin{figure*}
\centering
\includegraphics[scale=0.78]{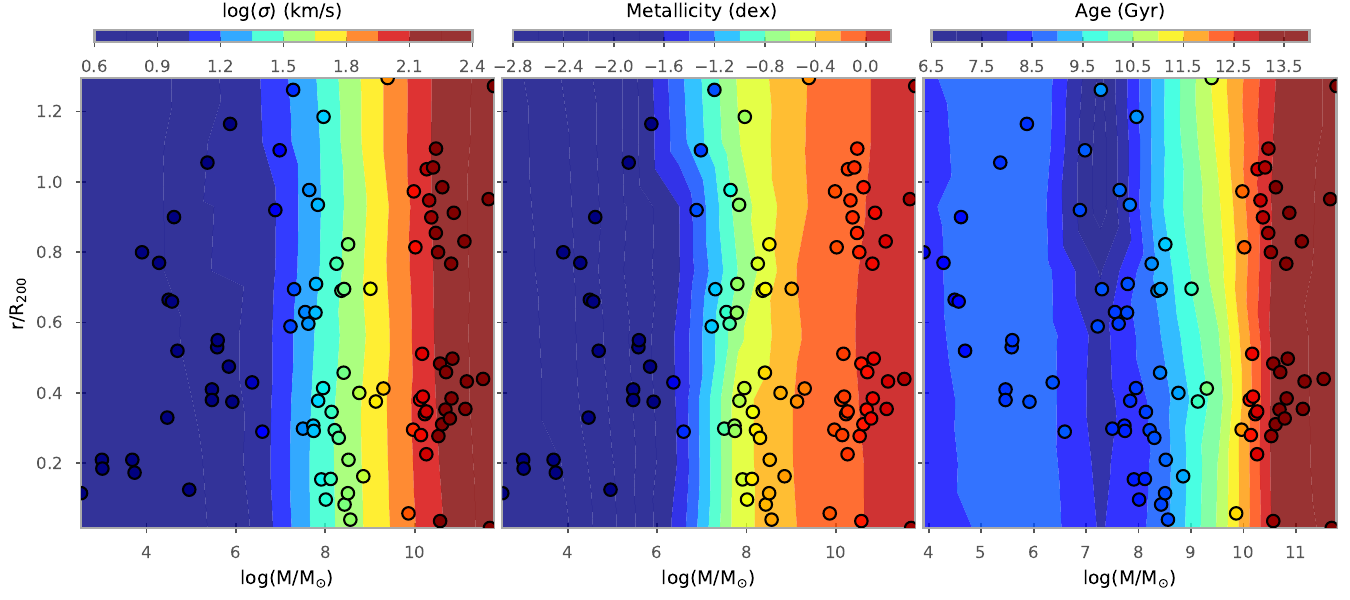}
\caption{Color map of the different properties of the galaxies as a function of their position in the projected distance-stellar mass plane. The panels show from left to right the logarithmic velocity dispersion, metallicity, and age. The scatter points are smoothed using the LOESS technique, and then with \href{https://docs.scipy.org/doc/scipy/reference/generated/scipy.spatial.KDTree.html}{Scipy.KDTree} the background maps are extrapolated. The almost vertical lines that are defined by the maps, tell us that these quantities are independent of the environment and that they are mainly driven by internal properties like the stellar mass.}
\label{fig_maps_vertical}
\end{figure*}
To explore this possible connection between stellar mass, environment and the formation time of galaxies, we show in Fig. \ref{fig_maps_times} the projected distances of the galaxies as a function of their stellar mass, and as a 3rd dimension we coloured them according to different properties. While on Fig. \ref{fig_2Dcolor_map_mass} we interpolated a continuous set of arrays to create the maps, now we want to create the map from a not uniformly distributed scatter of points, so in order to create the 2D maps of the properties we first smooth the points using the LOESS technique \citep{LOESS_2D}.
This algorithm estimates a regression fit for each point using nearby points to create a smooth curve. The technique is used when working with noisy data or scatter plots in order to smooth the results, with the intention of revealing underlying trends or patterns. Once the distribution of points has been smoothed, we used again the \href{https://docs.scipy.org/doc/scipy/reference/generated/scipy.spatial.KDTree.html}{Scipy.KDTree} to produce a continuous 2D map.

In Fig. \ref{fig_maps_times} we show the distribution of the galaxies in the projected distance - stellar mass plane, with each galaxy coloured according to their formation times t$_{20}$, t$_{50}$, t$_{70}$ and t$_{90}$. As a background, Fig. \ref{fig_maps_times} also shows the extrapolated 2D map of the different formation times of the galaxies in the \local, \sami and \atlas samples. We see that for galaxies more massive than 10$^{10}$M$_{\odot}$ the environment does not seem to have any significant effect on the evolution of the galaxy at any point in time, which is represented by the vertical iso-contours. It has to be kept in mind that these results only apply to the present-day environment, and some of these relatively massive galaxies (\sml10$^{10}$M$_{\odot}$) could have been preprocessed in smaller structures that were later accreted into the final cluster. In general, this means that the galaxy downsizing holds for this mass regime.

In general, at the moment that the galaxies have formed 20\% of their current mass, the environment has not done anything notable yet and all galaxies reached this point $\sim$13 Gyr ago. This could indicate that at the beginning of the formation of the galaxy, the galaxies were mostly isolated and were not affected by the environment yet. Later in time, when the galaxies have reached 50\% of their mass, even though the scale is small, we detect a more significant trend as a function of the environment for galaxies up to 10$^{8}$M$_{\odot}$. At this point, the galaxies in the inner regions have already started to experience some environmental quenching, and the maps begin to exhibit a U-shape. The parabolic shape becomes even more obvious for the t$_{70}$ and t$_{90}$ maps in Fig. \ref{fig_maps_times}. Here, the times of massive galaxies are always independent of the environment, and for less massive galaxies only those with a stellar mass around 10$^{8}$M$_{\odot}$ have kept forming stars until recently. For even less massive dwarfs like the ones in our \local, there is almost no environmental effect at this point, because the gas has already been blown away independently of its stellar mass.

We want to remark here that our sample is not homogeneously distributed, and there are several areas in Fig. \ref{fig_maps_times} that do not contain many galaxies. As a result, the extrapolation for those under mapped regions could bias our interpretations. However, judging from the behaviour of the \sami dwarfs, we expect that less massive dwarfs in less dense environments, like a group, are quenched as fast as indicated by Fig. \ref{fig_maps_times}.

\subsection{Environment independent properties}
We have discussed how the SFHs relate to both internal and external properties, but not all properties depend on both factors. Figure \ref{fig_maps_vertical} shows the location of each galaxy in the projected cluster distance - stellar mass plane, colored by: velocity dispersion, metallicity, and age.

\begin{figure*}
\centering
\includegraphics[scale=0.78]{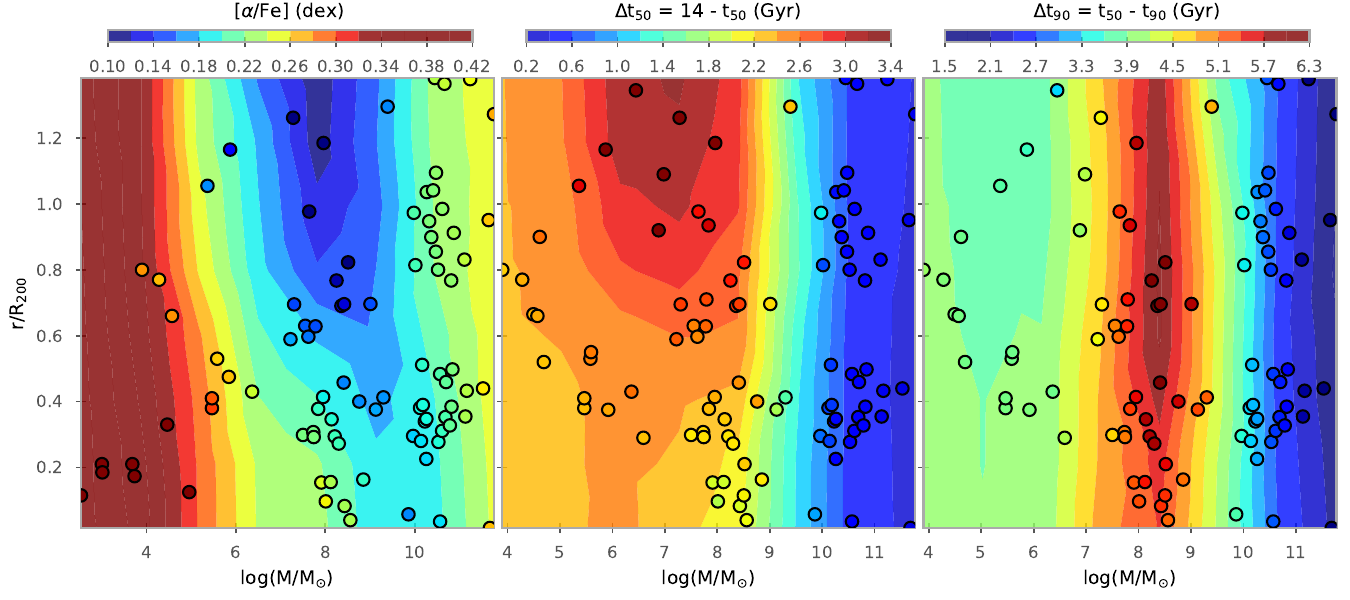}
\caption{Color map of the different properties of the galaxies as a function of their position in the projected distance-stellar mass plane. The panels show from left to right the \alfe, $\Delta$t$_{50}$ and $\Delta$t$_{90}$. The scatter points are smoothed using the LOESS technique, and then with \href{https://docs.scipy.org/doc/scipy/reference/generated/scipy.spatial.KDTree.html}{Scipy.KDTree} the background maps are extrapolated. For all the panels we can see traces of the U-shape that dictates the evolution time scales.}
\label{fig_maps_u_shape}
\end{figure*}
Our analysis reveals that there appears to be no correlation between the age, metallicity, or velocity dispersion of a galaxy and its location within a cluster or group of galaxies. These properties are primarily dependent on the internal properties of the galaxy, like its stellar mass, as represented by the vertical iso-contours shown in Fig. \ref{fig_maps_vertical}.

The velocity dispersion is determined by the dynamical mass, and part of that is the stellar mass. The relation has been extensively studied for early-type galaxies \citep{Auger2010_vel_disp}, at different redshifts \citep{Cannarozzo2020_vel_disp_z}, and even with analogous quantities such as luminosity \citep{Faber_Jackson_1976, Eftekhari_2021_fornaxII}. 

The metallicity-stellar mass relation is usually explained by self enrichment in the potential wells of the galaxies \citep{Dekel_Silk_1986}: the more massive a galaxy is the more metals from the stellar and supernova winds it can retain, which are then reprocessed in subsequent star formation events. Although it is also related with the star formation rate \citep{Mannucci2010}. The relation is maintained for all the masses \citep{Kirby2013} and morphological types \citep{Calura2009_mass_metal_morph}. Although some studies such as \citet{Ellison_2009_mass_metal_cluster} and \citet{Thomas2005} have found some that galaxies in high-density environments could be slightly more metal-rich (by about 0.05 dex), they conclude that these differences are related to local density changes \citepalias{Romero-Gomez_2022}.

Like the SFHs, age is related to the galaxy mass. Previous studies have already stated that dwarfs are on average younger than giants \citep{Sybilska2017, Romero-Gomez_2022}. As for the environment, \citet{Mateo2008} study the Local Group and found that most dSph have intermediate-old ages and are close to the host. To the contrary, young galaxies like dTrans and dIrr can be found mostly beyond the virial radius. Given that we do not analyze star-forming galaxies here, we can only agree on the ages of dSph being independent of the distance to the host.

\subsection{Formation time scales}\label{discus_timescales}

Our results show that less massive dwarfs are dominated by environmental processes, resulting in the SFHs of these types of galaxies being peaked and similar to those of giants galaxies. The similarities have been already pointed out in the past. \citet{Leitner2012} compared photometry-based SFHs of nearby star forming galaxies, the lowest average stellar mass was 10$^{7.6}$M$_{\odot}$, with spectra-based SFHs of massive galaxies from the SDSS, which reached a stellar mass up to 10$^{11.3}$M$_{\odot}$. Studying the star formation rate-stellar mass relation, they noticed that the least massive galaxies have SFHs that resembled those of the highest mass galaxies. This was inconsistent with the downsizing trend obtained from the SDSS sample, and they discussed that this could be due to technical problems derived from systematic errors in the models rather than a change in the physical reasons. We also acknowledge that systematic effects could be introduced by the different methodologies used to obtain the SFHs. However, this change in the stellar mass relations that we present as a U-shape, has been detected not only in the SFHs, but also in the [$\alpha$-Fe] values and even in simulations. \citet{Akins2021-simulations} studied simulations of dwarfs satellites in Milky Way-like halos and found that for most galaxies with masses between 10$^{6}$ and 10$^{8}$M$_{\odot}$ the environment produced a fast quenching. This agrees well with Fig. \ref{fig_maps_times}, where we see that Local Group galaxies are quenched in a few Gyr while more massive dwarfs keep forming stars.

To shed more light on this issue, we use again the projected distance-stellar mass plane. The SFHs, as already noticed by \citet{Leitner2012}, could be affected by some systematics derived from the models. For that reason, we now look at parameters that represent how fast the star formation happened, independently of when it starts. The first of these parameters is the [$\alpha$/Fe], which we already studied in \citetalias{Romero-Gomez_2022}. The $\alpha$-elements, like Mg or C among others, are produced in Supernovae type II, while Fe is mostly produced in Supernovae type Ia. The firsts are the final stage of the evolution of stars that have lived very short lives, while the latter comes from progenitors with a very long lifespan. Thus, the [$\alpha$-Fe] values could be used as a proxy of the timescales of star formation in a galaxy \citep{Reynier1989PhD, Worthey1992}. We can obtain the other timescale parameters from the formation times we derived from the cumulative SFHs \citep{Ferre-Mateu2018}. Using 14 Gyr as an approximation for the time of the Big Bang \citep{Planck2014}, we can define $\Delta$t$_{50}$ = 14 - t$_{50}$. This scale is representative of how long it took for the galaxy to build up 50\% of its mass, or similarly the mass of the halo \citep{Tojeiro2017_t50_halo}. Then we can also define $\Delta$t$_{90}$ = t$_{50}$ - t$_{90}$. With these three parameters, we cover how fast the overall star formation was, [$\alpha$/Fe], and then how fast the galaxy was in forming the first half of its mass, $\Delta$t$_{50}$, and then the second half, $\Delta$t$_{90}$.

In Figure \ref{fig_maps_u_shape} we present again the location of the galaxies in the projected cluster distance -stellar mass plane coloured this time by their \alen, $\Delta$t$_{50}$ and $\Delta$t$_{90}$.
In all three panels, we can see the U-shape effect, meaning that a given property linearly relates to stellar mass until 10$^{7}$-10$^{8}$M$_{\odot}$, and then the slope of that relation starts to be the opposite. 
The result is that, as we have discussed before, the lowest massive dwarfs have $\Delta$t$_{90}$ closer to that of the giants than those of intermediate-mass dwarfs.
Despite the similarity, this does not mean that the physical processes responsible for the evolution of dwarfs and giants are the same \citep{Haines2007}.

Looking at the $\Delta$t$_{50}$ and $\Delta$t$_{90}$ of giant galaxies in the central and right panels of Fig. \ref{fig_maps_u_shape},  these galaxies were all quenched 2-3 Gyr after the Big Bang, which means that probably most of them were already quenched before falling to their cluster. In any case, we see that they self-regulate their own evolution, as shown by the vertical iso-contours of Fig. \ref{fig_maps_u_shape}. Dwarf galaxies, on the other hand, are expected to be quenched by the environment \citep{Barmentloo2023}. In the central panel of Fig. \ref{fig_maps_u_shape} we see how the first half of the SFH is completed faster by those dwarfs that are close to their host. For the second half of the SFH only galaxies with stellar masses between 10$^{7}$ and 10$^{9}$M$_{\odot}$ have kept forming stars until recently. This mass range keeps appearing as a turning point in galaxy downsizing, which could be interpreted as an indication that only certain processes occur above/below this threshold. \citet{Sawala2010} studied the effects of Supernovae type II and Ia, as well as the effect of UV background radiation on a sample of simulated dwarf galaxies. They found that galaxies with total mass below $\sim$10$^{8}$M$_{\odot}$ could be quenched by the combination of these two processes. The majority of the dwarfs studied here do not have t$_{90}$ compatible with being quenched at the end of the re-ionization epoch (z$\sim$6 or 12.5 Gyr \citep{Gallart2013_reionization}), which point out to the environment being most likely responsible for the quenching. However, some studies have found that the least massive dwarfs could also be quenched only because of supernovae feedback \citep{Gallart2021_sn_feeedback}.

Looking at the whole picture of our results, we see that they provide strong evidence supporting the idea that the evolution of dwarf galaxies, with \superscript[4] < \stelmas < \superscript[8] \solmas, is dictated by their stellar mass and their local environment \citep{Thomas2010}. The TNG50-simulations of dwarfs in the mass regime of our \sami studied by \citet{Joshi2021} agree well with our assumptions. They concluded that the SFHs were strongly dependent on the stellar mass, with the less massive being assembled earlier, and the t$_{90}$ relations flattening for stellar masses between 10$^{9}$ and 10$^{10}$M$_{\odot}$. They also found that the environment was a secondary dependence, that could be negligible in the case of central galaxies or those with M$_{\star}$ > 10$^{10}$M$_{\odot}$. Gathering all the information it seems that for the less massive dwarfs, M$_{\star}$ < 10$^{7}$M$_{\odot}$, the environment is responsible for blowing most of their gas away. The little amount of gas remaining is then consumed in a fast burst of star formation, which produces enough supernovae feedback to completely quench the galaxy. As a result of this process, the [$\alpha$/Fe] of these galaxies remains high. More massive dwarfs, between 10$^{8}$ and 10$^{9}$M$_{\odot}$, endure the environmental process and are able to keep enough gas to keep forming stars until recently. Even more massive galaxies, the giants, as have been extensively studied in the literature, are barely affected by various environments.

As pointed out before, in future works it would be interesting to see if the star-forming dwarfs deviate from the previous relations and if they follow the continuity of galaxy downsizing at lower masses \citep{Kolova2011}. In \citet{Romero-Gomez_2022} we highlight a small sample of dwarfs from the \sami survey, and a few from the Local Group, that followed a relation more similar to the one proposed by the downsizing. Probably one of the main reasons behind this behaviour is that these young systems are mostly outside the virial radius of the host or are infalling at present, which indicates that there has not been enough time for them to be quenched by the environment.
\section{Summary and Conclusions}\label{conclusions}
We present the analysis of the star formation histories of 31 dwarfs elliptical galaxies in the Fornax cluster, obtained with full spectral fitting algorithms. For comparison, using the same methodology, we obtained the SFHs from the 260 ETGs galaxies of the ATLAS$^{3D}$ project. At the low mass end, we include a sample of photometric SFHs of dwarf galaxies from the \local. After a careful inspection of the data we draw the following conclusions:

\begin{itemize}
  \item The shape of the SFHs is strongly related to the stellar mass of the galaxy. Massive galaxies formed fast, in a few Gyr after the Big Bang, while the star-formation rate for less massive galaxies is smaller. This galaxy downsizing continues until dwarfs of M$_{\star}$ $\sim$10$^{7}$-10$^{8}$M$_{\odot}$. Dwarf galaxies with smaller stellar masses, on the other hand, quenched earlier.
  \item We find that the environment does not influence the evolution of massive galaxies much. For bright dwarfs, there is a combination between internal and external effects, while for faint dwarfs with stellar mass lower than \superscript[6] \solmas the evolution is mostly determined by the environment.
  \item We considered the location of galaxies in a plane of projected distance vs. stellar mass to inspect the dependence of formation time scales on external and internal factors. Galaxies more massive than 10$^{10}$M$_{\odot}$ are completely independent of the environment and evolve according to their own internal properties. For less massive galaxies we find that the time scales are distributed following a U-shape. This means that down to $\sim$10$^{8}$M$_{\odot}$ dwarfs follow the downsizing scheme that comes from the giants, while for less massive galaxies we find an 'upsizing', giving them an SFH shape similar to that of the giants.
  \item Other parameters like the velocity dispersion or the metallicity appear to be independent of the environment, and their values are only related to internal properties. The age seems also to be affected only by the stellar mass, although not linearly because given the U-shape of the time scales the less massive dwarfs from the \local are older than those of our \sami sample.
\end{itemize}

In summary, only galaxies with stellar masses between 10$^{8}$ and 10$^{9}$M$_{\odot}$ can resist the quenching induced by different environmental processes, retaining enough gas to keep forming stars until recently. Giants are barely affected by the environment, and the least massive galaxies can lose their gas as soon as they arrive at a high-density environment. This behaviour agrees well with simulations, suggesting that 10$^{8}$M$_{\odot}$ represents a threshold in galaxy evolution. For future investigations, it would be interesting to see if this U-shape relation can be confirmed in other galaxy clusters, groups and other types of galaxies, as well as for other important properties. These findings could have significant implications for our understanding of the complex relation of SFHs with galaxy properties and pave the way for future investigations in this exciting field.

\section*{Data availability}
The reduced data underlying this article will be made available through the CDS. The raw data is publicly available in the AAT data archive.

\section*{Acknowledgements}
JRG and JALA are supported by the Spanish Ministry of Education, Culture and Sports under grant AYA2017-83204-P and by
the Spanish Ministerio de Ciencia e Innovaci\'on y Universidades by the grant PID2020-119342GB-I00.

For the analysis we have used Python \href{http://www.python.org}{http://www.python.org}; Matplotlib \citep{Hunter2007}, a suite of open source python modules that provide a framework for creating scientific plots; and Astropy, a community-developed core Python package for Astronomy \citep{Astropy2013}.



\bibliographystyle{mnras}
\bibliography{References} 




\appendix

\section{Star formation histories of \atlas galaxies}\label{appendix_atlas}
To expand the sample of this work, we combined our dwarf sample with the giant galaxies of ATLAS$^{3D}$ \citep[][]{Cappellari2011-atlas3d}. As explained in \citet{Romero-Gomez_2022}, we collapsed all spatially resolved spectra into one single spectrum per galaxy, and using the same methodology as for the dwarfs we obtained the stellar populations parameters and SFH.

\begin{figure*}
  \centering
  \subfloat[]{\includegraphics[scale=0.9]{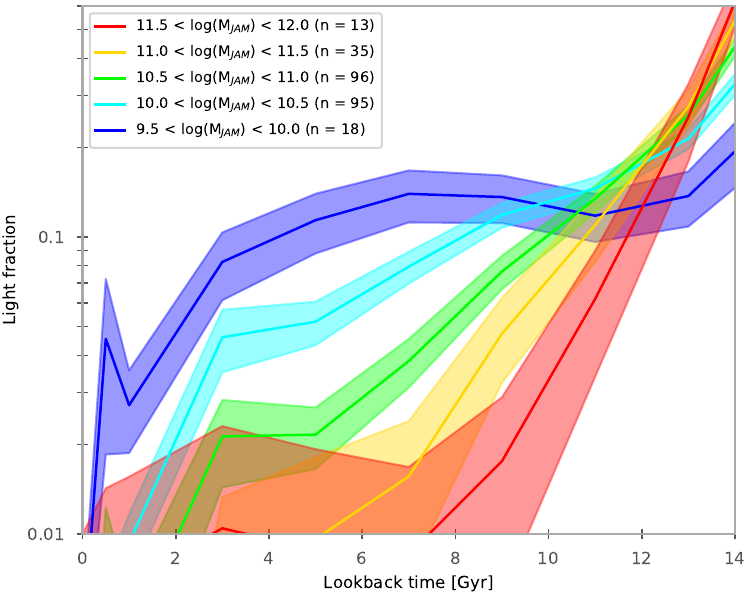}\label{fig:f1}}
  \hfill
  \subfloat[]{\includegraphics[scale=0.33]{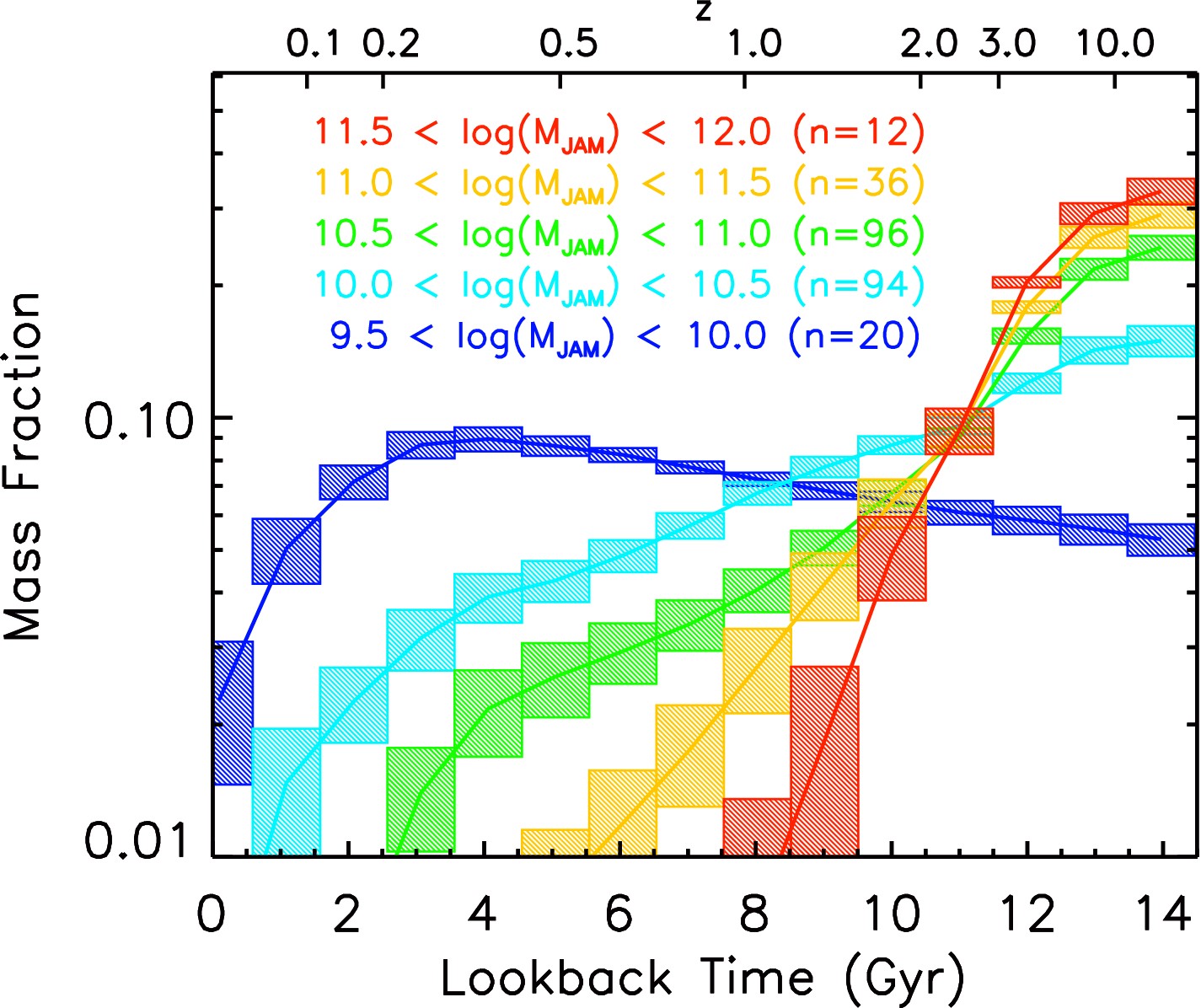}\label{fig:f2}}
  \caption{In the top panel, (a), we show the SFH of the \atlas galaxies derived in our study, while in the bottom panel are the SFH of the \atlas galaxies derived by \citet{McDermid2015}. For a better comparison of both results we binned the galaxies using the same mass bin as in \citetalias{McDermid2015}. The color of each bin is stated in the legend of the figures. In both cases is clear that the more massive galaxies formed faster in a few Gyr after the Big Bang.}
  \label{fig_compare_mass_fraction_mcdermid}
\end{figure*}

\begin{figure*}
\centering
\includegraphics[scale=0.7]{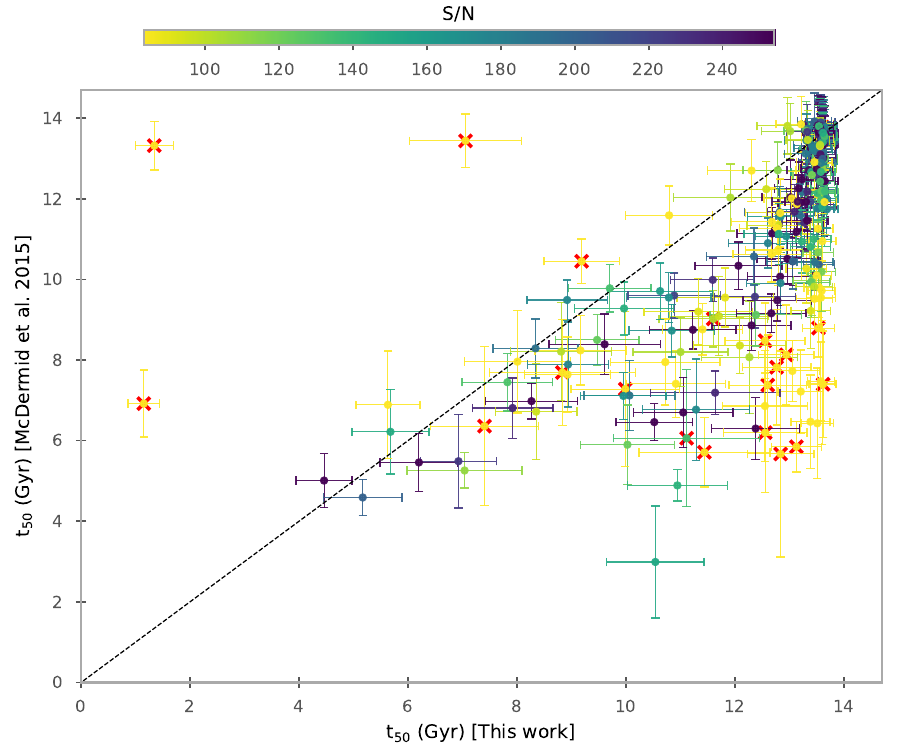}
\caption{Comparison of our t$_{50}$ with those published in \citetalias{McDermid2015} for the \atlas galaxies. Each galaxy is colour coded with the S/N, and the red X are those points of bad quality according to \citetalias{McDermid2015}. The 1:1 relation is represented by the black dashed line on each panel.}
\label{fig_compare_times_mcdermid_apx}
\end{figure*}

In Fig. \ref{fig_compare_mass_fraction_mcdermid} we show a comparison between the SFHs we derived, and those from \citet{McDermid2015}. For the sake of comparison, we combined the SFHs in the same stellar mass bins \citet{McDermid2015} used. As a grid for the diagram they used the models from \citet{Schiavon2007}, and for this reason, their values are not directly comparable to ours with different methodologies and template models. Despite the differences, we can see that the SFHs are fairly similar. Adiotionally, in Fig. \ref{fig_compare_times_mcdermid_apx} we also show a comparison of the t$_{50}$ form our work, and that of \citet{McDermid2015}. There we see some differences between both values, but these are expected given the different models and methodologies used, as explain in \citetalias{Romero-Gomez_2022}.

\section{Star formation histories of Local Group galaxies}\label{appendix_Local_weisz}

\begin{figure*}
\centering
\includegraphics[scale=0.7]{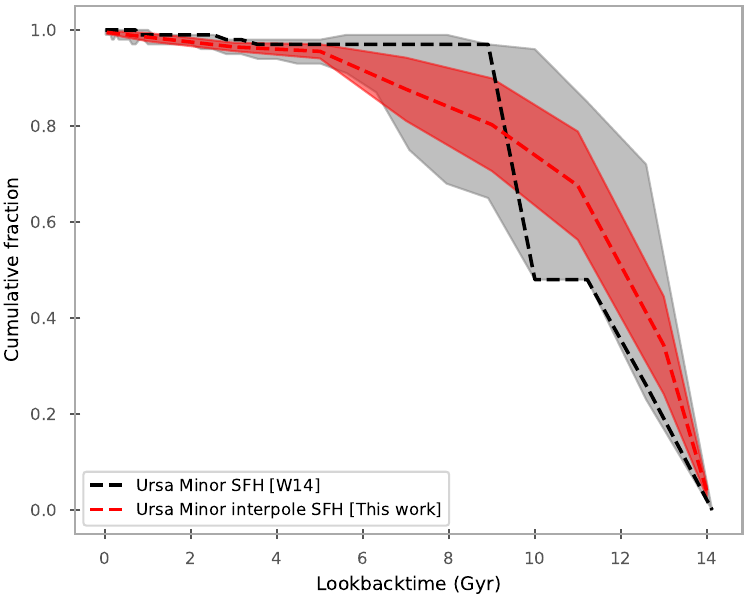}
\caption{Example of the interpolation on the SFH of the Ursa Minor galaxy from \citetalias{Weisz_2014_sfh}. The SFH and erros from \citetalias{Weisz_2014_sfh} are the dashed black line and grey shadow, respectively. The interpolated SFH and error we used in this work is represented by the red dash line and shadow, respectively.}
\label{fig_interpole_weisz_apx}
\end{figure*}

\begin{figure*}
\centering
\includegraphics[scale=0.7]{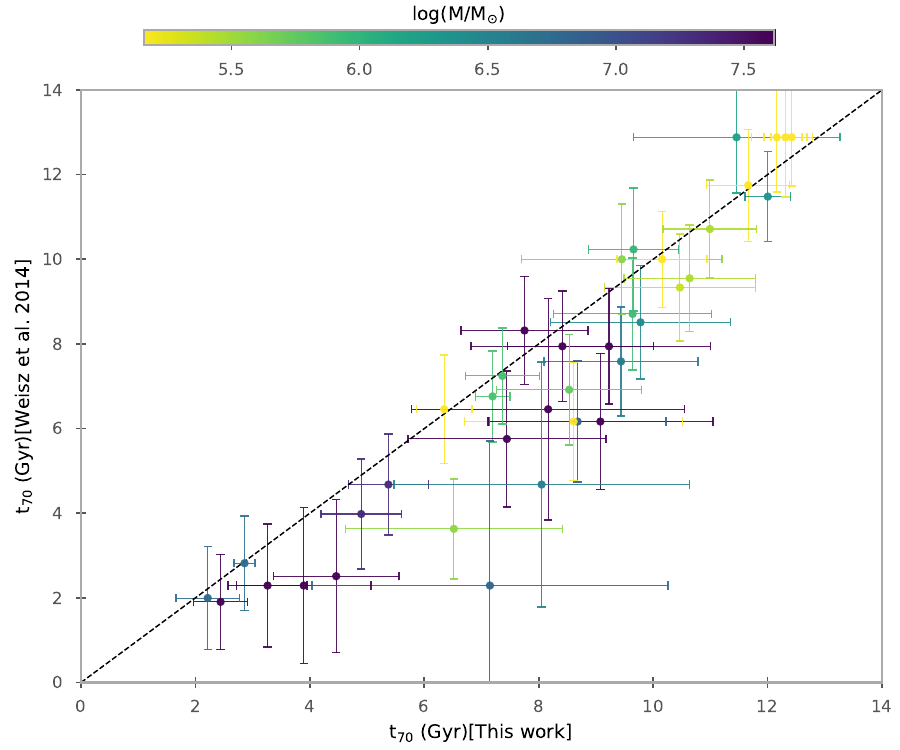}
\caption{Comparison of our t$_{70}$ with those published in \citetalias{Weisz_2014_sfh} for the \local galaxies. Each galaxy is colour coded with its stellar mass, and the 1:1 relation is represented by the black dashed line on each panel.}
\label{fig_compare_times_weisz_apx}
\end{figure*}

In order to have the SFHs from the \local in the same age grid as our \sami and \atlas results, we interpolated the SFHs from \citetalias{Weisz_2014_sfh}. We used their errors and Monte Carlo simulations to compute the errors of the interpolation, as explain in Section \ref{results_formation_times}. In Fig. \ref{fig_interpole_weisz_apx} we show a comparison of the SFH from \citetalias{Weisz_2014_sfh} for the Ursa Minor galaxy, and our interpolation. To test this, in Fig. \ref{fig_compare_times_weisz_apx} we also compare the t$_{70}$ published by \citet{Weisz_2014_t70} with the times computed from our interpolation method. In general there is a well agreement between both values.
\section{Star formation histories of the SAMI-Fornax dwarfs}\label{appendix_sami}
In this appendix we present tables with the information relative to the SFHs of the \sami dwarfs. Table \ref{table_apx_1} has for each galaxy the weight, corresponding to the formed light fraction, at different lookback times. Table \ref{table_apx_2} has the formation times used in this paper.
\begin{table*}
\caption{Table with the star formation histories of the SAMI-Fornax dwarfs. The first column shows the galaxy name, while the rest of the columns are the different lookback times in Gyr. Then for each galaxy the rows has the name and the light fraction formed at the corresponding lookback time.} 
\centering    
\begin{tabular}{ccccccccccc}     
\hline
\input{Tables/table_C1.txt}

\end{tabular}
             
\label{table_apx_1}
\end{table*}
\begin{table*}
\caption{Table with the formation times of the \sami dwarfs. The columns are: galaxy name, t$_{20}$, t$_{50}$, t$_{70}$, t$_{90}$ and the name of the corresponding data file on CDS. The data in CDS are the reduce data cubes. All the times are in lookback time Gyr. } 
\centering    
\begin{tabular}{cccccc}     
\hline
\input{Tables/table_C2.txt}

\end{tabular}
             
\label{table_apx_2}
\end{table*}

\bsp	
\label{lastpage}
\end{document}

%% file: Tables/table_C1.txt
FCC name & 0.04 & 0.5 & 1.0 & 3.0 & 5.0 & 7.0 & 9.0 & 11.0 & 13.0 & 14.0 \\

\noalign{\smallskip}
\hline\noalign{\smallskip}

\noalign{\smallskip}
FCC100 & 0.047 & 0.013 & 0.066 & 0.136 & 0.08 & 0.138 & 0.163 & 0.125 & 0.106 & 0.126 \\

\noalign{\smallskip}
FCC106 & 0.004 & 0.0 & 0.047 & 0.066 & 0.084 & 0.1 & 0.228 & 0.298 & 0.15 & 0.023 \\

\noalign{\smallskip}
FCC134 & 0.263 & 0.04 & 0.298 & 0.123 & 0.066 & 0.069 & 0.058 & 0.024 & 0.017 & 0.044 \\

\noalign{\smallskip}
FCC135 & 0.057 & 0.0 & 0.001 & 0.189 & 0.061 & 0.036 & 0.14 & 0.27 & 0.187 & 0.06 \\

\noalign{\smallskip}
FCC136 & 0.009 & 0.001 & 0.003 & 0.07 & 0.086 & 0.086 & 0.166 & 0.255 & 0.229 & 0.095 \\

\noalign{\smallskip}
FCC143 & 0.019 & 0.002 & 0.0 & 0.001 & 0.003 & 0.007 & 0.054 & 0.247 & 0.414 & 0.252 \\

\noalign{\smallskip}
FCC164 & 0.048 & 0.0 & 0.013 & 0.306 & 0.041 & 0.05 & 0.087 & 0.15 & 0.221 & 0.084 \\

\noalign{\smallskip}
FCC178 & 0.03 & 0.022 & 0.036 & 0.069 & 0.119 & 0.166 & 0.186 & 0.17 & 0.125 & 0.077 \\

\noalign{\smallskip}
FCC181 & 0.106 & 0.013 & 0.03 & 0.021 & 0.024 & 0.007 & 0.049 & 0.042 & 0.196 & 0.511 \\

\noalign{\smallskip}
FCC182 & 0.004 & 0.0 & 0.001 & 0.011 & 0.033 & 0.094 & 0.223 & 0.312 & 0.237 & 0.085 \\

\noalign{\smallskip}
FCC188 & 0.027 & 0.0 & 0.02 & 0.111 & 0.065 & 0.081 & 0.143 & 0.216 & 0.223 & 0.115 \\

\noalign{\smallskip}
FCC195 & 0.073 & 0.005 & 0.008 & 0.025 & 0.037 & 0.091 & 0.192 & 0.248 & 0.2 & 0.121 \\

\noalign{\smallskip}
FCC202 & 0.009 & 0.008 & 0.043 & 0.108 & 0.168 & 0.197 & 0.188 & 0.148 & 0.092 & 0.041 \\

\noalign{\smallskip}
FCC203 & 0.071 & 0.0 & 0.008 & 0.435 & 0.064 & 0.031 & 0.062 & 0.135 & 0.151 & 0.042 \\

\noalign{\smallskip}
FCC211 & 0.039 & 0.014 & 0.024 & 0.062 & 0.112 & 0.156 & 0.179 & 0.175 & 0.144 & 0.095 \\

\noalign{\smallskip}
FCC222 & 0.005 & 0.001 & 0.01 & 0.033 & 0.076 & 0.152 & 0.255 & 0.234 & 0.115 & 0.119 \\

\noalign{\smallskip}
FCC223 & 0.044 & 0.0 & 0.001 & 0.016 & 0.035 & 0.077 & 0.221 & 0.335 & 0.14 & 0.13 \\

\noalign{\smallskip}
FCC245 & 0.028 & 0.032 & 0.064 & 0.109 & 0.147 & 0.166 & 0.161 & 0.136 & 0.098 & 0.059 \\

\noalign{\smallskip}
FCC250 & 0.298 & 0.008 & 0.029 & 0.06 & 0.082 & 0.143 & 0.109 & 0.101 & 0.104 & 0.066 \\

\noalign{\smallskip}
FCC252 & 0.024 & 0.001 & 0.002 & 0.003 & 0.006 & 0.01 & 0.036 & 0.166 & 0.453 & 0.297 \\

\noalign{\smallskip}
FCC253 & 0.088 & 0.008 & 0.069 & 0.17 & 0.157 & 0.209 & 0.123 & 0.058 & 0.063 & 0.055 \\

\noalign{\smallskip}
FCC264 & 0.023 & 0.0 & 0.006 & 0.009 & 0.019 & 0.067 & 0.182 & 0.307 & 0.225 & 0.161 \\

\noalign{\smallskip}
FCC266 & 0.017 & 0.001 & 0.01 & 0.007 & 0.016 & 0.085 & 0.283 & 0.361 & 0.163 & 0.057 \\

\noalign{\smallskip}
FCC274 & 0.038 & 0.022 & 0.031 & 0.017 & 0.023 & 0.047 & 0.088 & 0.097 & 0.186 & 0.451 \\

\noalign{\smallskip}
FCC298 & 0.018 & 0.0 & 0.007 & 0.019 & 0.016 & 0.054 & 0.202 & 0.174 & 0.32 & 0.19 \\

\noalign{\smallskip}
FCC300 & 0.031 & 0.009 & 0.034 & 0.024 & 0.048 & 0.086 & 0.177 & 0.257 & 0.162 & 0.172 \\

\noalign{\smallskip}
FCC301 & 0.013 & 0.003 & 0.132 & 0.095 & 0.078 & 0.119 & 0.171 & 0.178 & 0.117 & 0.093 \\

\noalign{\smallskip}
FCC277 & 0.023 & 0.037 & 0.084 & 0.141 & 0.178 & 0.181 & 0.153 & 0.109 & 0.064 & 0.03 \\

\noalign{\smallskip}
FCCB442 & 0.181 & 0.316 & 0.296 & 0.12 & 0.027 & 0.014 & 0.01 & 0.008 & 0.009 & 0.018 \\

\noalign{\smallskip}
FCCB904 & 0.047 & 0.002 & 0.016 & 0.03 & 0.036 & 0.069 & 0.176 & 0.172 & 0.172 & 0.28 \\

\hline\noalign{\smallskip}

%% file: Tables/table_C2.txt
FCC name & t$_{20}$ (Gyr) & t$_{50}$ (Gyr) & t$_{70}$ (Gyr) & t$_{90}$ (Gyr) & File name \\

\noalign{\smallskip}
\hline\noalign{\smallskip}

\noalign{\smallskip}
FCC100 & 13.76$\pm$0.2 & 11.86$\pm$1.02 & 6.91$\pm$1.87 & 1.46$\pm$0.88 & fcc100.fits \\

\noalign{\smallskip}
FCC106 & 13.06$\pm$0.25 & 11.01$\pm$0.33 & 9.0$\pm$0.71 & 5.07$\pm$0.93 & fcc106.fits \\

\noalign{\smallskip}
FCC134 & 6.08$\pm$1.33 & 2.14$\pm$0.5 & 1.14$\pm$0.47 & 0.24$\pm$0.21 & fcc134.fits \\

\noalign{\smallskip}
FCC135 & 13.69$\pm$0.2 & 12.06$\pm$0.96 & 8.89$\pm$1.61 & 4.38$\pm$1.51 & fcc135.fits \\

\noalign{\smallskip}
FCC136 & 13.69$\pm$0.21 & 12.52$\pm$0.66 & 9.84$\pm$1.15 & 4.99$\pm$1.09 & fcc136.fits \\

\noalign{\smallskip}
FCC143 & 13.96$\pm$0.06 & 13.89$\pm$0.09 & 13.39$\pm$0.08 & 12.18$\pm$0.27 & fcc143.fits \\

\noalign{\smallskip}
FCC164 & 13.72$\pm$0.2 & 12.74$\pm$0.59 & 10.15$\pm$1.78 & 4.14$\pm$1.4 & fcc164.fits \\

\noalign{\smallskip}
FCC178 & 11.99$\pm$1.14 & 6.94$\pm$1.56 & 3.87$\pm$1.01 & 1.92$\pm$1.09 & fcc178.fits \\

\noalign{\smallskip}
FCC181 & 13.77$\pm$0.24 & 13.73$\pm$0.24 & 13.09$\pm$0.53 & 7.54$\pm$3.29 & fcc181.fits \\

\noalign{\smallskip}
FCC182 & 13.69$\pm$0.17 & 12.12$\pm$0.31 & 10.11$\pm$0.4 & 6.71$\pm$0.83 & fcc182.fits \\

\noalign{\smallskip}
FCC188 & 13.81$\pm$0.17 & 13.27$\pm$0.42 & 12.16$\pm$0.84 & 8.11$\pm$2.27 & fcc188.fits \\

\noalign{\smallskip}
FCC195 & 13.76$\pm$0.23 & 13.03$\pm$0.57 & 8.97$\pm$2.63 & 3.22$\pm$2.5 & fcc195.fits \\

\noalign{\smallskip}
FCC202 & 13.49$\pm$0.29 & 10.38$\pm$0.84 & 8.53$\pm$0.64 & 5.6$\pm$1.04 & fcc202.fits \\

\noalign{\smallskip}
FCC203 & 13.7$\pm$0.2 & 12.66$\pm$0.64 & 8.23$\pm$1.9 & 3.41$\pm$1.12 & fcc203.fits \\

\noalign{\smallskip}
FCC211 & 13.25$\pm$0.5 & 11.38$\pm$1.05 & 9.55$\pm$1.02 & 6.94$\pm$1.4 & fcc211.fits \\

\noalign{\smallskip}
FCC222 & 13.31$\pm$0.45 & 11.12$\pm$0.56 & 9.73$\pm$0.54 & 6.83$\pm$1.3 & fcc222.fits \\

\noalign{\smallskip}
FCC223 & 13.55$\pm$0.3 & 11.96$\pm$1.04 & 10.77$\pm$1.46 & 6.98$\pm$1.55 & fcc223.fits \\

\noalign{\smallskip}
FCC245 & 13.05$\pm$0.47 & 7.69$\pm$1.55 & 3.89$\pm$0.93 & 2.04$\pm$0.82 & fcc245.fits \\

\noalign{\smallskip}
FCC250 & 11.24$\pm$1.62 & 6.98$\pm$1.02 & 4.69$\pm$1.57 & 1.93$\pm$1.16 & fcc250.fits \\

\noalign{\smallskip}
FCC252 & 13.95$\pm$0.07 & 13.89$\pm$0.09 & 13.53$\pm$0.13 & 12.59$\pm$0.4 & fcc252.fits \\

\noalign{\smallskip}
FCC253 & 13.75$\pm$0.24 & 12.88$\pm$0.5 & 10.12$\pm$1.79 & 4.34$\pm$1.88 & fcc253.fits \\

\noalign{\smallskip}
FCC264 & 13.79$\pm$0.21 & 13.63$\pm$0.24 & 12.44$\pm$0.76 & 9.18$\pm$2.31 & fcc264.fits \\

\noalign{\smallskip}
FCC266 & 12.03$\pm$0.47 & 10.33$\pm$0.38 & 9.13$\pm$0.55 & 6.86$\pm$1.16 & fcc266.fits \\

\noalign{\smallskip}
FCC274 & 13.84$\pm$0.17 & 13.5$\pm$0.3 & 11.08$\pm$1.38 & 4.11$\pm$2.44 & fcc274.fits \\

\noalign{\smallskip}
FCC298 & 13.46$\pm$0.34 & 11.26$\pm$0.8 & 9.34$\pm$0.88 & 7.19$\pm$1.23 & fcc298.fits \\

\noalign{\smallskip}
FCC300 & 13.45$\pm$0.36 & 11.23$\pm$0.76 & 8.35$\pm$1.21 & 3.43$\pm$1.62 & fcc300.fits \\

\noalign{\smallskip}
FCC301 & 12.85$\pm$0.24 & 9.76$\pm$0.24 & 6.91$\pm$0.53 & 2.9$\pm$0.39 & fcc301.fits \\

\noalign{\smallskip}
FCC277 & 8.86$\pm$0.26 & 6.37$\pm$0.18 & 4.9$\pm$0.26 & 2.46$\pm$0.53 & fcc277.fits \\

\noalign{\smallskip}
FCCB442 & 2.92$\pm$0.3 & 1.75$\pm$0.3 & 1.18$\pm$0.31 & 0.51$\pm$0.31 & fccB442.fits \\

\noalign{\smallskip}
FCCB904 & 13.61$\pm$0.25 & 11.97$\pm$1.1 & 9.46$\pm$1.2 & 4.72$\pm$2.18 & fccB904.fits \\

\hline\noalign{\smallskip}